\documentclass{aa}

\usepackage{graphicx}

\usepackage{txfonts}
\usepackage{siunitx}
\DeclareSIUnit\gauss{G}
\usepackage{ragged2e}
\usepackage{natbib}
\usepackage{subfigure}
\usepackage{multirow}

\usepackage{hyperref}
\hypersetup{
    colorlinks=true,
    linkcolor=blue,
    citecolor=blue,
    filecolor=magenta,      
    urlcolor=cyan,
}

\begin{document}

\title{The role of ambipolar heating in the energy balance of solar prominences}

   \author{Lloren\c{c} Melis
          \inst{1,2}
          \and
          Roberto Soler\inst{1,2}}

   \institute{Departament de Física, Universitat de les Illes Balears, E-07122, Palma de Mallorca, Spain\\ \and Institut d'Aplicacions Computacionals de Codi Comunitari (IAC3), Universitat de les Illes Balears, E-07122, Palma de Mallorca, Spain}

\titlerunning{Ambipolar heating in solar prominences}
 \authorrunning{Melis \& Soler}
 
\abstract{Solar prominence threads are typically located around magnetic dips, where cold and dense plasma is suspended against gravity in the hot corona thanks to the upward magnetic force. Because prominences are partially ionized, ambipolar diffusion can deposit part of the energy of their non-force-free magnetic field into the plasma. This ambipolar heating may therefore play a role in the energy balance of prominences. In this proof-of-concept work, we explore the effect of ambipolar diffusion in one-dimensional  models that satisfy both mechanical equilibrium and energy balance. The magnetic configuration is based on the classic Kippenhahn–Schl\"uter model, incorporating a sheared magnetic field. The temperature profile along the magnetic field is computed numerically by balancing radiative losses, thermal conduction, and ambipolar heating. The resulting models consistently consist of a cold, dense, partially ionized thread with prominence core conditions, a very thin prominence-corona transition region, and an extended, hot, fully ionized region with coronal conditions. In addition to providing heating that partly compensates for radiative losses, ambipolar diffusion also gives rise to stationary flows associated with the gravitational drainage of neutrals in the partially ionized region. We investigate how the length of the cold threads depends on the central temperature, central pressure, magnetic field strength, and shear angle, and show that thread lengths compatible with observations are obtained for realistic values of these parameters. Therefore, we demonstrate that ambipolar diffusion plays a relevant role in this simple configuration, indicating that this effect should be incorporated into more elaborate multi-dimensional models and simulations.}


\maketitle
\section{Introduction}

Solar prominences are one of the most intriguing objects in the solar atmosphere. They consist of masses of relatively cool and dense  plasma suspended in the corona. Although living in the corona,  their plasma properties are more similar to  those in the chromosphere \citep[see, e.g.][]{vial2015prominences}. An important property of the prominence plasma is that it is partially ionized
\citep[see, e.g.,][]{heinzel2024}. Prominences are composed of numerous long and thin ribbons, called threads, which are only seen in high-resolution observations \citep[see, e.g.,][]{lin2011filaments}. These threads presumably outline  the prominence magnetic structure, whose field lines are anchored at the photosphere \citep[see, e.g.][]{martin2015prominences}. The threads are believed to reside in magnetic dips, which allow the dense prominence plasma to be supported against gravity  \citep[e.g.,][]{anzer2007}, although this idea has been challenged \citep{Karpen2001}.

Explaining how prominences and their threads are formed in the coronal medium is a complicated topic and several different mechanisms may be involved \citep[see][]{karpen2015,Zhou2025}. However, there is an increasing consensus  that the key process that allows the condensation of the cool prominence plasma is related to thermal instability \citep[for a recent review, see][and references therein]{keppens2025}. In this context, the role of partial ionization in the formation through the evaporation-condensation process has been investigated by \citet{jercic2025}. Once formed,  prominences can remain in a relative stable global state for long periods of time before their eventual disappearance \citep[see][]{parenti2014prominences}.  The present paper does not deal with the formation of prominences. Instead, we focus on studying their equilibrium.

There is a rich literature on prominence equilibrium models. The mechanical equilibrium  is a problem well understood, as it is  established that the upward force produced by the magnetic field is required to compensate gravity \citep[see][]{gibson2018}. A pioneering work in this regard is the classic paper by \cite{kippenhahn1957filament}, hereafter referred as the K-S model, where a magnetohydrostatic equilibrium is constructed. In the K-S model, the weight of the prominence plasma produces a shallow dip in a nearly horizontal magnetic field. Although many subsequent papers are based on the K-S model, it represents the most simplified approach and more elaborate models exist in the literature \citep[see, e.g.,][to name a few]{Kuperus1974,ballester1987model,hood1990,heinzel2001prominence,low2005structure,Blokland2011}. Some applications of the K-S model include the study of prominence oscillations \citep[e.g.,][]{oliver1992waves,oliver1993oscillations,anzer2009} and instabilities  \citep{hillier2010magnetic}.

Contrary to the mechanical equilibrium, the energy balance is a problem  not completely solved \citep[see, e.g.,][]{gilbert2015balance}. Although such a balance must necessarily result from the interplay between cooling, heating, and thermal conduction, their relative importance and, particularly, the nature of the heating remain under study \citep[see][]{anzer1999balance}. As prominences are suspended in the corona, illumination from the photosphere and the surrounding corona is an important heating source \citep[see][]{poland1971energy,heasley1976strucutre,Heinzel2025}. However, radiative equilibrium models obtained by solving the full non-local thermodynamic equilibrium (non-LTE)  radiative-transfer problem typically produce slightly lower equilibrium temperatures than those expected in prominence cores \citep[see, e.g.][]{heinzel2010prominences,heinzel2012radiative}. Recently, \citet{gunar2025radiative} computed net radiative cooling rates from 1D non-LTE isothermal and isobaric models \citep[see also][]{Heinzel2025}. These net rates result from the combined effects of  radiative cooling and radiative heating. According to  \citet{gunar2025radiative}, a net loss of energy occurs for typical conditions in prominence cores.  Their results suggest that, although radiative heating is most likely the dominant heating mechanism in prominences, another complementary heating could be necessary to completely balance the net energy losses. 

Several heating mechanisms may be acting in the coronal medium \citep[see, e.g.,][]{mandrini2000currents,arregui2024}. In the context of prominences, some works have used parametrized forms of the heating aiming to represent different processes in a simplified  way without explicitly solving the underlying physics \citep[see, e.g.,][]{brughmans2022rope}. A potential heating source is related to the dissipation of magnetohydrodynamics (MHD) waves \citep[e.g.,][]{pecseli2000,parenti2007,soler2016heat,melis2021heat,Hashimoto2023}. Since the prominence plasma is partially ionized, another possibility is the dissipation of currents by ambipolar diffusion. Such heating mechanism has been investigated in the chromosphere \citep[e.g.,][]{goodman2004,khomenko2012heating,sykora2017,Stepanov2024}. However, its importance in the energy balance of prominences has not been examined yet.

Prominence equilibrium models that simultaneously satisfy both mechanical equilibrium and energy balance have  been explored in the literature using different approaches \citep[e.g.,][to name a few]{poland1971energy,low1975,milne1979prominences,low1981,degenhardt1993flux,schmitt1995,vigh2018model}. Our purpose is to construct such kind of models, but including the effect of ambipolar diffusion. To this end, we deliberately adopt a simple approach based on the use of a modified K-S configuration, as way to test the importance of ambipolar heating in the energy balance.  The currents associated with the non-force-free magnetic field of the K-S model are susceptible to be dissipated by ambipolar diffusion, which would cause the deposition of magnetic energy into the plasma. Such heating would affect the energy balance and the relative importance of radiative losses and thermal conduction.

The method we use to construct the models is based on those  of \citet{terradas2021thread} and \citet{melis2023heat}, which is restricted to finding one-dimensional equilibria that satisfy the energy balance equation along a magnetic field line. In \citet{terradas2021thread} an arbitrarily imposed uniform heating was used and the magnetic field was prescribed instead of being  computed from the balance of forces.  \cite{melis2023heat} extended the method by considering a non-uniform heating resulting from the dissipation of Alfv\'en waves. \cite{melis2023heat} implemented a numerical scheme in which the Alfv\'en wave equation and the energy balance equation were solved iteratively until convergence to a self-consistent equilibrium was achieved. However, \cite{melis2023heat} also prescribed the magnetic field.  Here, we follow an iterative strategy as in \cite{melis2023heat}, but with two important differences. On the one hand, we include ambipolar diffusion as a heating mechanism instead of wave heating. On the other hand, we  couple the balance of forces with the energy equation. As a result, the obtained models satisfy both mechanical equilibrium and energy balance, and include ambipolar diffusion in a consistent manner.

This paper is structured as follows. Section~\ref{sec:method} includes an explanation of the method and the self-consistent approach. The results of the computed models are presented in Sect.~\ref{sec:results}. Finally, some general conclusions are given in Sect. \ref{sec:conclusions}. Additionally, Appendix~\ref{sec:resolution} discusses the convergence of the numerical method used to compute the  models and Appendix~\ref{sec:beta} includes a comparison of our results with those of the related work by \citet{milne1979prominences}.

\section{Method}
\label{sec:method}

\subsection{Basic equations for partially ionized plasma}

In this work we used the single-fluid MHD equations for partially ionized plasmas \citep[e.g.,][]{ballester2018partial}. The basic equations are,
\begin{eqnarray}
    \frac{D \rho}{Dt} &=& -\rho \nabla \cdot \vec{v}, \label{eq:continuity} \\
    \rho \frac{D\vec{v}}{Dt} &=& -\nabla p + \frac{1}{\mu_{0}} \left(\nabla \times \vec{B}\right) \times \vec{B}+\rho \vec{g}, \label{eq:momentum} \\
    \frac{\partial \vec{B}}{\partial t} &=& \nabla \times \left( \vec{v} \times \vec{B} \right) +\nabla \times \Bigl\{ \eta_{\rm  A} \left[ \left(\nabla \times \vec{B}\right) \times \vec{B} \right]\times \vec{B} \Bigr\}, \label{eq:induction} \\
    \frac{Dp}{Dt} &=& -\gamma p \nabla \cdot \vec{v} + (\gamma-1)\mathcal{L}, \label{eq:energy}\\
    \nabla \cdot \vec{B}&=&0, \label{eq:gauss}
\end{eqnarray}
with $\frac{D }{Dt} = \frac{\partial }{\partial t} + \vec{v}\cdot \nabla$ the material derivative. These equations are, respectively, the continuity equation, the momentum equation, the induction equation, the energy equation, and the solenoidal condition of the magnetic field. The symbols in these equations have their usual meaning: $\vec{v}$ is the velocity, $p$ is the gas pressure, $\rho$ is the mass density, $\mu_{0}$ is the magnetic permeability, $\vec{B}$ is the magnetic field, $\vec{g}$ is the acceleration of gravity, $\eta_{\rm A}$ is the ambipolar diffusion coefficient, $\gamma$ is the adiabatic index, and $\mathcal{L}$ the heat-loss function, which encloses all sources and sinks of energy, such as radiative cooling, thermal condition, and ambipolar heating. This is elaborated later.

The relation between pressure and temperature, $T$, can be written with the ideal gas law, namely
\begin{equation}
    p = N k_{\rm B}T,
    \label{eq:gasn}
\end{equation}
where $N$ is the total number density and $k_{\rm B}$ is Boltzmann's constant. We considered a partially ionized plasma composed of hydrogen and helium. The total number density is computed as
\begin{equation}
    N = \sum_{\beta} n_{\beta},
\end{equation}
where $n_{\beta}$ denotes the number density of the different species $\beta$ of the plasma. The species considered are electrons (e), protons (p), neutral hydrogen (H), neutral helium (HeI), singly ionized helium (HeII) and doubly ionized helium (HeIII).  The number densities of the various species can be computed as functions of total mass density, $\rho$, as
\begin{eqnarray}
    n_{\rm p} &=& \frac{1}{1+4A}\frac{\xi_{\rm p}\,\rho}{m_{\rm p}}, \\
    n_{\rm H} &=& \frac{1}{1+4A}\frac{\xi_{\rm H}\,\rho}{m_{\rm p}}, \\
    n_{\rm He\,j} &=& \frac{A}{1+4A}\frac{\xi_{\rm He\,j}\,\rho}{4m_{\rm p}},
\end{eqnarray}
for $j=$~I, II, and III, where $A$ is the helium to hydrogen abundance ratio, $m_{p}$ is the proton   mass, and $\xi_\beta$ denotes the mass fraction of the corresponding  species $\beta$. In turn, owing to charge conservation, the electron number density satisfies,
\begin{equation}
    n_{\rm e} = n_{\rm p} +n_{\rm HeII} + n_{\rm HeIII}. \label{eq:ne}
\end{equation}
We considered  $A=0.1$, which means an abundance of helium  of 10$\%$ with respect to that of hydrogen. Therefore, $N$ can be reformulated as
\begin{equation}
    N = \frac{1.1+\xi_{\rm p}+0.1\xi_{\rm HeII}+0.2\xi_{\rm HeIII}}{1.4m_{\rm p}}\rho.
\end{equation}
These definitions allow us to rewrite Eq. (\ref{eq:gasn}) in the usual form,
\begin{equation}
    p = \frac{\rho R T}{\tilde{\mu}},
    \label{eq:gas}
\end{equation}
where $R=k_{\rm B}/m_{\rm p}$ is the ideal gas constant and $\tilde{\mu}$ is the mean atomic weight, which in our case is written down as
\begin{equation}
   \tilde{\mu} = \frac{1.4}{1.1+\xi_{\rm p}+0.1\xi_{\rm HeII}+0.2\xi_{\rm HeIII}}.
   \label{eq:weight}
\end{equation}

The mass fractions of the various species can be used to denote their ionization degree. Hence, $\xi_{\rm p}$ is used to denote the hydrogen ionization, while $\xi_{\rm HeII}$ and $\xi_{\rm HeIII}$ denote the two different states of helium ionization. In order to set $\xi_{\rm p}$,  we used the values in  Table~1 from \cite{heinzel2015fast} for an altitude of 20,000~km above the photosphere, which are obtained from non-LTE computations in 1D prominence slabs\footnote{Updated values of the hydrogen ionization fraction are provided in \cite{gunar2025radiative}. However, the differences with the previous results are small and have a negligible impact on our results.}. The values of the table were interpolated and for $T \geq 2 \times 10^{4}$ K we considered full ionization of hydrogen, i.e.,  $\xi_{\rm p}=1$. However, to our knowledge, no similar non-LTE computations are currently available  in the case of helium. To circumvent this limitation, we adopted a simplified LTE approach using the Saha equation. The expressions of  $\xi_{\rm HeII}$ and $\xi_{\rm HeIII}$ are,
\begin{eqnarray}
\xi_{\rm HeII} = \frac{\mathcal{N}_{\rm HeII}}{1+\mathcal{N}_{\rm HeII}+\mathcal{N}_{\rm HeII} \mathcal{N}_{\rm HeIII}}, \\
\xi_{\rm HeIII} = \frac{\mathcal{N}_{\rm HeII} \mathcal{N}_{\rm HeIII}}{1+\mathcal{N}_{\rm HeII}+\mathcal{N}_{\rm HeII} \mathcal{N}_{\rm HeIII}},
\end{eqnarray}
where $\mathcal{N}_{\rm HeII}$ and $\mathcal{N}_{\rm HeIII}$ represent the ratio between the number density in the corresponding ionization state and that in the previous ionization stage. The ratios can be computed as
\begin{eqnarray}
    \mathcal{N}_{\rm HeII} = \frac{2}{n_{\rm e}}\frac{Z_{\rm HeII}}{Z_{\rm HeI}}\left( \frac{2\pi m_{\rm e}k_{\rm B}T}{h^{2}} \right)^{3/2}e^{-\frac{\chi_{\rm HeI}}{k_{\rm B}T}}, \\
    \mathcal{N}_{\rm HeIII} = \frac{2}{n_{\rm e}}\frac{Z_{\rm HeIII}}{Z_{\rm HeII}}\left( \frac{2\pi m_{\rm e}k_{\rm B}T}{h^{2}} \right)^{3/2}e^{-\frac{\chi_{\rm HeII}}{k_{\rm B}T}},
\end{eqnarray}
where $Z_{\rm HeI} = Z_{\rm HeIII} =1 $ and $Z_{\rm HeII}=2$ are the partition functions of the ionization stages, $h$ is  Planck's constant, $m_{\rm e}$ is the electron mass, and $\chi_{\rm HeI} = 24.6$~eV and $\chi_{\rm HeII} = 54.4$~eV are the corresponding ionization energies.

\begin{figure}
    \centering
    \includegraphics[width=0.9\linewidth]{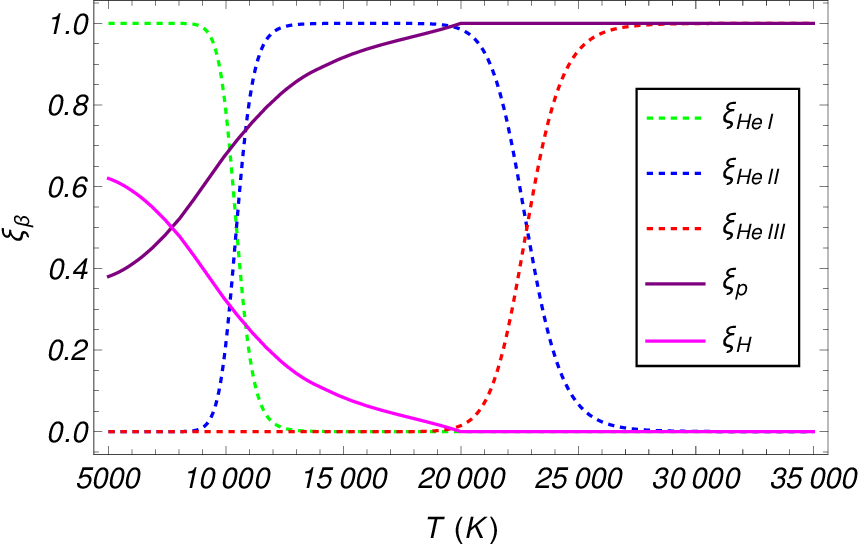}
    \caption{Ionization fractions of hydrogen and helium for $p=5\times10^{-3}$ Pa as functions of the temperature, $T$.}
    \label{fig:ionisation}
\end{figure}

The ionization fractions of helium in both ionization states are coupled with the hydrogen ionization fraction through the electron number density (Eq.~(\ref{eq:ne})). We solved the coupled system and computed the ionization fractions of both ionized helium species as functions of the temperature and pressure. Figure \ref{fig:ionisation} illustrates how the ionization fractions of all the considered species  vary as functions of $T$ for a prescribed value of $p$. For low temperatures, hydrogen is partially ionized, while helium is neutral. As $T$ increases, the progressive ionization of hydrogen follows the non-LTE results of  \cite{heinzel2015fast}, whereas the ionization of helium follows the LTE result from the Saha equation. For $T \approx$~11,000~K singly ionized helium becomes the dominant helium state, while for $T \approx$~23,000~K doubly ionized helium becomes dominant. Full ionization of helium is reached at $T \approx$~30,000~K.

We  considered in Eq.~(\ref{eq:induction}) the non-ideal  ambipolar diffusion, which is related with ion-neutral collisions, but  other  effects such  Ohmic diffusion, Hall's term, and the battery term are dropped for simplicity \citep[see, e.g.,][]{Khomenko2014,ballester2018partial}. Previous results in \citet{melis2023heat} indicate that ambipolar diffusion is the dominant effect in prominence conditions. The ambipolar diffusion coefficient, $\eta_{\rm A}$, in a partially ionized hydrogen-helium plasma  takes the expression \citep[see][]{Zaqarashvili2013,soler2015}:
\begin{equation}
    \eta_{\rm A} = \frac{\xi_{\rm H}^{2}\alpha_{\rm HeI}+\xi_{\rm HeI}^{2}\alpha_{\rm H}+2\xi_{\rm H}\xi_{\rm HeI}\alpha_{\rm H HeI}}{\mu_{0} \left( \alpha_{\rm HeI}\alpha_{\rm H}-\alpha_{\rm H HeI}^{2}\right)}, \label{eq:etaa}
\end{equation}
where $\alpha_{\beta}$ is the total friction coefficient of species $\beta$ (accounting for collisions of this particular species with all the others) and $\alpha_{\beta\beta'} = \alpha_{\beta'\beta}$ is the symmetric friction coefficient for collisions between species $\beta$ and $\beta'$. The total friction coefficients are computed as
\begin{equation}
    \alpha_{\beta} = \sum_{\beta \neq \beta'} \alpha_{\beta\beta'}.
\end{equation}
If both species are charged, the friction coefficient is \citep[e.g.,][]{Spitzer1962,Braginskii1965}:
\begin{equation}
    \alpha_{\beta \beta'} = \frac{n_{\beta}n_{\beta'}e^{4}\ln \Lambda_{\beta \beta'}}{6\pi\sqrt{2\pi}\epsilon_{0}^{2}m_{\beta \beta'}\left( k_{\rm B}T/m_{\beta\beta'}\right)},
\end{equation}
where  $m_{\beta\beta'}=m_{\beta}m_{\beta'}/(m_{\beta}+m_{\beta'})$ is the reduced mass, $e$ is the electron charge, $\epsilon_{0}$ is the electrical permittivity, and $\Lambda_{\beta\beta'}$ is the Coulomb logarithm, which is expressed as
\begin{equation}
    \ln \Lambda_{\beta \beta'} = \ln \left( \frac{24\pi\epsilon_{0}^{3/2}k_{\rm B}^{3/2}T^{3/2}}{e^{3}\sqrt{n_{\beta}+n_{\beta'}}} \right).
\end{equation}
If at least one of the colliding species is neutral, the friction coefficient is the approximation of small drift velocity becomes \citep[e.g.,][]{Braginskii1965,Chapman1970,Draine1986}:
\begin{equation}
    \alpha_{\beta \rm n} = n_{\beta}n_{\beta'}\sqrt{\frac{8k_{\rm B}T}{\pi m_{\beta \rm n}}} \sigma_{\beta \rm n},
\end{equation}
where $\sigma_{\beta n}$ is the collisional cross-section, whose values in the context of the solar atmosphere can be found in, e.g., \citet{vranjes2013} and \citet{Wargnier2022}.

\subsection{Mechanical equilibrium}

We used a Cartesian coordinate system. We assumed that gravity points in the negative $z$-direction, so that $\vec{g}=-g\hat{z}$, with $g$  the surface acceleration of gravity assumed constant. Seeking for an  equilibrium, we imposed $\partial/\partial t = 0$ in the MHD Eqs.~(\ref{eq:continuity})--(\ref{eq:energy}). Additionally, we assumed that all variables only vary along the $x$-direction. Under these conditions, the problem of finding an equilibrium is analogous to the classic K-S configuration, here modified by the presence of ambipolar diffusion owing to partial ionization. The effect of gravity should produce a curved magnetic field, i.e., a dip, where the prominence material can be supported. Therefore, we found that the horizontal components of the magnetic field, $B_x$ and $B_y$, are both constant, so that these two components can be expressed in terms of a uniform horizontal magnetic field strength,  $B_{0}$, and a shear angle $\phi = \arctan \left( B_{y}/B_{x}\right)$, namely
\begin{eqnarray}
    B_{x} &=& B_{0}\cos \phi, \\
    B_{y} &=& B_{0}\sin \phi.
\end{eqnarray}    
In turn, the vertical component of the magnetic field, $B_z$, follows the typical K-S form, namely
\begin{equation}
     B_{z} = \sqrt{2\mu_{0}p_{0}}\tanh \left( \frac{\sqrt{2\mu_{0}p_{0}}}{2B_{x}}\frac{g}{R} f(x) \right), \label{eq:magnetic}
\end{equation}
where $p_{0}$ is the pressure at the center of the magnetic dip (located at $x=0$) and the function $f(x)$ is \citep[see][]{poland1971energy},
\begin{equation}
    f(x) = \int_0^x \frac{\tilde{\mu}(x_1)}{T(x_1)}dx_1.
\end{equation}
Note that, at the current stage, $\tilde{\mu}$ and $T$ are still unknown, so that $f(x)$ cannot be explicitly computed yet. The pressure and density again follow the  K-S solution, namely 
\begin{eqnarray}
    p &=& p_{0} -\frac{B_{z}^{2}}{2\mu_{0}},\label{eq:pressure}\\
    \rho &=& \frac{B_{x}}{\mu_{0}g}\frac{d B_{z}}{dx}. \label{eq:density}
\end{eqnarray}
The density is  non-uniform  owing to the accumulation of the prominence mass  around the dipped part of the field line. In addition,  a non-uniform pressure is also required to preserve the force balance along the magnetic field line. 

While there are no equilibrium flows in the classic K-S equilibrium, i.e., it is static, such flows  appear when ambipolar diffusion is included, so that the equilibrium is stationary  \citep[see also][]{Bakhareva1992}. Assuming $v_x = 0$, the $y$- and $z$-components of the equilibrium velocity  are,
\begin{eqnarray}
    v_{y} &=& -\eta_{\rm A}\frac{B_{y}}{2 B_{x}}\frac{d B_{z}^{2}}{dx},\label{eq:vertical}\\
    v_{z} &=& -\eta_{\rm A} \frac{B_{x}^{2}+B_{z}^{2}}{B_{x}}\frac{d B_{z}}{dx}. \label{eq:drainage}
\end{eqnarray}
The result that $v_y$ and $v_z$ are both proportional to $\eta_{\rm A}$ plainly shows that these equilibrium flows are driven by ambipolar diffusion. The role of these flows, specially that of the vertical one, is explored later. We note that the general case with $v_x \ne 0$ is considerably more complicated \citep[see, e.g.,][]{tsinganos1992}. The main reason for adopting $v_x=0$ is that we seek equilibrium solutions that are essentially modified versions of the classic K-S model. If $v_x \ne 0$, terms proportional to $v_x$ and its spatial derivative appear in all the governing equations,  introducing effects such as the enthalpy flux in the energy balance equation. As such kind of siphon flows are obtained in time-dependent simulations \citep[see, e.g.,][]{xia2011,donne2024}, relaxing the condition $v_x=0$ would be an interesting extension of this study that may lead to the computation of a different family of solutions. This is left for future works.

\subsection{Energy balance}

The equilibrium obtained in the previous Subsection is still incomplete. The modified K-S solution depends on the function $f(x)$ that, in turn, requires $\tilde\mu$ and $T$ to be known. The mean atomic weight can be computed from Eq.~(\ref{eq:weight}). However, knowledge of the ionization state of the plasma is needed, which depends on the temperature. Therefore, computing $f(x)$ ultimately reduces to computing the temperature profile. To this end, we  solved the energy balance condition.

The heat-loss function, $\mathcal{L}$, on the right-hand side of Eq.~(\ref{eq:energy}) generally takes the expression,
\begin{equation}
   \mathcal{L} =  Q_{\rm A} -\nabla\cdot \vec{q}-L_{\rm rad.}+C, \label{eq:mathcalL}
\end{equation}
where $Q_{\rm A}$ is the ambipolar heating, $\vec{q} = -\hat\kappa\nabla T$ is the heat flux due to thermal conduction, with $\hat\kappa$  the thermal conductivity tensor, $L_{\rm rad.}$ represents the energy losses due to radiative cooling, and $C$ may represent other heating sources, such as viscous and/or Ohmic heating, which are here ignored.

The ambipolar heating  is calculated using the Joule heating function that accounts for the dissipation of perpendicular electric currents \citep[e.g.,][]{khomenko2012heating}, namely
\begin{equation}
    Q_{\rm A} = \mu_{0}|\vec{B}|^{2}\eta_{\rm A}|\vec{J}_{\perp}|^{2},
\end{equation}
where $\vec{J}_{\perp}$ is the perpendicular component of the current density with respect to the magnetic field direction. Since the equilibrium magnetic field in the K-S solution is not force free, there is necessarily  a nonzero current in the equilibrium. We have that
\begin{equation}
    \vec{J}_{\perp} = \frac{1}{\mu_{0}|\vec{B}|^{2}}\left( B_{y}B_{x}\hat{x}-(B_{x}^{2}+B_{z}^{2})\hat{y}+B_{z}B_{y}\hat{z} \right)\frac{dB_{z}}{dx}.
\end{equation}
After some algebra, we easily arrive at an expression for the heating rate, namely
\begin{equation}
    Q_{\rm A} = \frac{\eta_{\rm A}\left( B_{0}^{2}\cos^2\phi + B_{z}^{2} \right)}{\mu_{0}}\left( \frac{dB_{z}}{dx}\right)^2. \label{eq:heatingrate}
\end{equation}
Therefore, ambipolar diffusion of the magnetic field continuously provides a heating input to the plasma. For an equilibrium to exist, this heating needs to be compensated by thermal conduction and radiative cooling.

Concerning thermal conduction in a partially ionized plasma, the conductivity of charges is dominated by the component parallel to the magnetic field, while that of neutrals is essentially isotropic \citep[see, e.g.,][]{soler2015}. Consequently,  the divergence of the heat flux in our configuration reduces to,
\begin{equation}
    \nabla\cdot \vec{q} = - \frac{d }{d x} \left( \kappa_{\rm eff.} \frac{d T}{d x} \right),
\end{equation}
with the effective thermal conductivity, $\kappa_{\rm eff.}$, being approximated as,
\begin{equation}
    \kappa_{\rm eff.} \approx \frac{B_{x}}{\vert \vec{B}\vert}\kappa_{\rm \parallel,c}+\kappa_{\rm n},
\end{equation}
where $\kappa_{\rm \parallel,c}$ and $\kappa_{\rm n}$ are the parallel component of the conductivity of charges and the isotropic neutral conductivity, respectively. The role of the charges thermal conductivity is influenced by orientation of the magnetic field through the factor $B_x/\vert \vec{B}\vert$, since only the parallel component is relevant. These conductivities are computing by adding the contributions of the various species, namely
\begin{eqnarray}
    \kappa_{\rm \parallel,c} &=& \kappa_{\rm \parallel,e}+\kappa_{\rm \parallel,p}+\kappa_{\rm \parallel,HeII}+\kappa_{\rm \parallel,HeIII}, \\
    \kappa_{\rm n} &=& \kappa_{\rm H}+\kappa_{\rm HeI},
\end{eqnarray}
where the species-related conductivities are computed as,
\begin{equation}
    \kappa_{\parallel,\beta} = 3.2 \frac{n_{\beta}^{2}k_{\rm B}^{2}T}{\alpha_{\beta}+\alpha_{\beta\beta}},
\end{equation}
for $\beta = $~e, p, HeII, and HeIII, and
\begin{equation}
  \kappa_{\beta} = \frac{5}{3}\frac{n_{\beta}^{2}k_{\rm B}^{2}T}{\alpha_{\beta}+\alpha_{\beta\beta}},
\end{equation}
for $\beta = $~H and HeI. In these expressions, $\alpha_{\beta\beta}$ denotes the  friction coefficient for internal self-collisions of species $\beta$, while all the other quantities have been defined before.

Now, we discuss the treatment of radiative losses. In principle, the net losses should be computed  from the combined effects of radiative cooling and radiative heating, as done recently by \citet{gunar2025radiative} using non-LTE calculations. The net losses tabulated in \citet{gunar2025radiative} depend on the location of the plasma element with respect to the source of illumination, i.e., the depth within the prominence.  However, the numerical procedure used here to generate models, based on that from \citet{melis2023heat}, requires the cooling rate to be expressed as a function of the local plasma properties alone. The implementation of the tables from \citet{gunar2025radiative} requires a  modification of  the numerical scheme of \citet{melis2023heat} that is beyond the aims of the present paper and is left for a forthcoming work. Therefore,  we use a simplified treatment to approximate the net radiative losses.  The form of the radiative cooling function we use is,
\begin{equation}
    L_{\rm rad.} = n_{\rm e}(n_{\rm p}+n_{\rm H}) \Lambda(T),
\end{equation}
This expression can be rewritten in terms of the total mass density as,
\begin{equation}
    L_{\rm rad.} = \frac{\xi_{\rm p}+0.1\xi_{\rm HeII}+0.2\xi_{\rm HeIII}}{1.4^{2}m_{p}^{2}}\rho^{2}\Lambda(T), \label{eq:radiation}
\end{equation}
where $\Lambda(T)$ in this and the previous expression is a piecewise function that depends on the temperature. Here we use the SPEX\_DM curve from \citet{hermans2021cooling}\footnote{We used an analytical fit of the numerically computed cooling curve available at \url{https://erc-prominent.github.io/team/jorishermans/}}. This is, essentially, an optically-thin cooling curve modified with a special treatment for low temperatures ($T<10^4$~K), as described in \citet{schure2009spex}. This and similar curves have recently been used in prominence simulations \citep[see, e.g.,][among others]{brughmans2022rope,jercic2025}. As explained by \citet{Heinzel2025}, the non-LTE net cooling rates may deviate significantly from  optically-thin calculations, so that caution is needed on this matter.

The condition of energy balance is formally,
\begin{equation}
    \mathcal{L} = 0,  \label{eq:balance}
\end{equation}
which results in a complicated second-order nonlinear differential equation for  $T$ that needs to be solved numerically. The full expression is omitted here for the sake of simplicity. We note that, apart from the fundamental constants, all physical quantities are spatially dependent in the model, with the exception of the horizontal component of the magnetic field and the shear angle.

\subsection{Numerical solution and self-consistent strategy}

We considered a single magnetic field line of length $L$. We used the coordinate $s$ to denote length along the magnetic field line, with the  center located at $s=0$ and the ends at $s=\pm L/2$. The parametric equations for the magnetic field lines are,
\begin{equation}
    \frac{dx}{B_{x}} = \frac{dy}{B_{x}} = \frac{dz}{B_{z}} = \frac{ds}{|\vec{B}|}.
\end{equation}
If the horizontal component dominates, as expected, then $B_{z} \ll B_{0}$ and  $|\vec{B}| \approx B_{0}$. Therefore, the coordinates $x$ and $y$ are almost directly proportional to $s$, since  $B_{x}$ and $B_{x}$ are both uniform. The parametric representation of the magnetic field line is, 
\begin{eqnarray}
    x &\approx & x_{0}+s\cos{\phi}, \\
    y &\approx & y_{0}+s\sin{\phi}, \\
    z &\approx & z_{0} + \int_0^s \frac{B_{z}(s_1)}{B_{0}} ds_1, \label{eq:z}
\end{eqnarray}
where $x_{0}$, $y_{0}$ and $z_{0}$ are arbitrary constants, so that $x_{0}=y_{0}=z_{0}=0$ can be imposed with no loss of generality. Note that the actual height of the field line is irrelevant since the K-S model is invariant in the vertical direction.

We used Wolfram Mathematica to numerically solve the energy balance condition (Eq.~(\ref{eq:balance})). The integration was performed with the module \verb|NDSolve|, which adapts the spatial resolution in order to minimize the numerical error. The numerical integration  was performed in two stages: first from $x=0$ to $x=x_{\rm max.}/2$, and then from $x=0$ to $x=-x_{\rm max.}/2$, where $x_{\rm max.}$ is the maximum value of the $x$ coordinate that corresponds with $s=L/2$.  The two solutions were joined together and the complete profile was constructed. In all cases, we considered a total length of $L=100$~Mm.

We imposed the same boundary conditions as in  \citet{terradas2021thread} and \citet{melis2023heat}. We prescribed a central temperature, $T_{0}$, and required that the temperature to be minimum at the center, namely
\begin{eqnarray}
    T=T_{0} \quad \textrm{at} \quad x=0, \\
    \frac{dT}{dx}=0 \quad \textrm{at} \quad x=0.
\end{eqnarray}
In addition, \citet{terradas2021thread}  showed that an additional restriction must be satisfied at the center for an equilibrium to exist under such conditions, namely
\begin{equation}
    \frac{d^2T}{dx^2} = \frac{L_{\rm rad.}-Q_{\rm A}}{\kappa_{\rm eff.}} >0, \quad \textrm{at} \quad x=0,
    \label{eq:constraint}
\end{equation}
so that $L_{\rm rad.}>Q_{\rm A}$ at the center. If $Q_{\rm A} > L_{\rm rad.}$, the heating would be too intense, making it impossible the existence of an equilibrium.

 Equilibrium solutions were studied considering four  parameters: the central pressure, $p_{0}$, the  horizontal magnetic field strength, $B_{0}$, the shear angle, $\phi$, and the central temperature, $T_{0}$. The considered ranges of values for those parameters are: from $2\times 10^{-3}~{\rm Pa}< p_0 <1.5\times10^{-2}~{\rm Pa}$, $4~{\rm G}< B_{0}< 20~{\rm G}$, $0^{\circ} < \phi < 90^{\circ}$, and 7,000~K~$<T_{0}<$~10,000~K. The central temperatures and pressures were chosen to be consistent with those expected in prominence cores, while the  magnetic field strength corresponded to that in quiescent prominences \citep[see, e.g.][]{gibson2018}. Importantly, these four parameters are not completely independent from each other regarding the existence of an equilibrium solution, so that it is not possible to compute  equilibria for certain combinations of parameters. This is explored later.

In a similar fashion as in \citet{melis2023heat} for the case of Alfv\'en wave heating, the computation of equilibrium  models that incorporated  ambipolar heating  posed a circular problem. The heating rate depends on the magnetic field, temperature, and other quantities through Eq.~(\ref{eq:heatingrate}). However, the temperature is obtained by solving the energy balance condition (Eq.~\ref{eq:balance}), which needs the heating rate to be known in advance. We solved this circular problem with an iterative strategy. First, we selected particular values of $B_{0}$, $p_{0}$, $\phi$ and $T_{0}$ for which the self-consistent  model is to be computed. The iterative method as initialized by solving Eq.~(\ref{eq:balance}) with no ambipolar heating ($Q_{\rm A}=0$). An initial temperature profile was obtained in such a way, which allowed the computation of the vertical magnetic field (Eq.~(\ref{eq:magnetic})), the pressure (Eq.~(\ref{eq:pressure})), and the density (Eq.~(\ref{eq:density})) profiles. With these initial profiles, the ambipolar heating rate was computed (Eq.~(\ref{eq:heatingrate})), completing the first iteration. Then, in the second iteration, the energy balance condition is again solved, but now using the ambipolar heating rate computed in the previous iteration, and the profiles for temperature, vertical magnetic field, pressure, and density are updated, which allow the computation of another ambipolar heating rate, and so on. 

The profiles obtained after every iteration would  be slightly different compared to those of the previous iteration. To quantify the difference from iteration to iteration, we used the $L^{2}$ relative error norm of the temperature profile, since it is the one computed from the energy balance condition. The error estimation, $\varepsilon$, is computed with the $L^{2}$ error norm as,
\begin{equation}
    \varepsilon = \sqrt{\frac{\int_{-x_{\rm max.}}^{x_{\rm max.}}\left( T_{i}(x)-T_{i-1}(x)\right)^{2} dx}{\int_{-x_{\rm max.}}^{x_{\rm max.}} T_{i-1}^{2}(x)\, dx}}, \label{eq:epsilon}
\end{equation}
where $T_{i}(x)$ and $T_{i-1}(x)$ are the temperature profiles for the $i$-th iteration  and the previous iteration, respectively. We assumed that a self-consistent equilibrium  model is achieved when $\varepsilon < \varepsilon_{0}$, where $\varepsilon_{0}$ is a prescribed small tolerance. We used $\varepsilon_{0}=10^{-7}$. The iterative process was repeated until the convergence criterion is satisfied. As mentioned before, for some combination of parameters the constraint of Eq.~(\ref{eq:constraint}) cannot be enforced. This causes the convergence criterion to be unachievable even after many iterations. In those cases,  it was not possible to compute an equilibrium model. The relevance of the spatial resolution and the effect of the constraint of Equation (\ref{eq:constraint}) are explored in Appendix~\ref{sec:resolution}.

\section{Results}
\label{sec:results}
\subsection{Properties of a reference  model}

\label{sec:typ}

\begin{figure}
    \centering
    \includegraphics[width=0.99\linewidth]{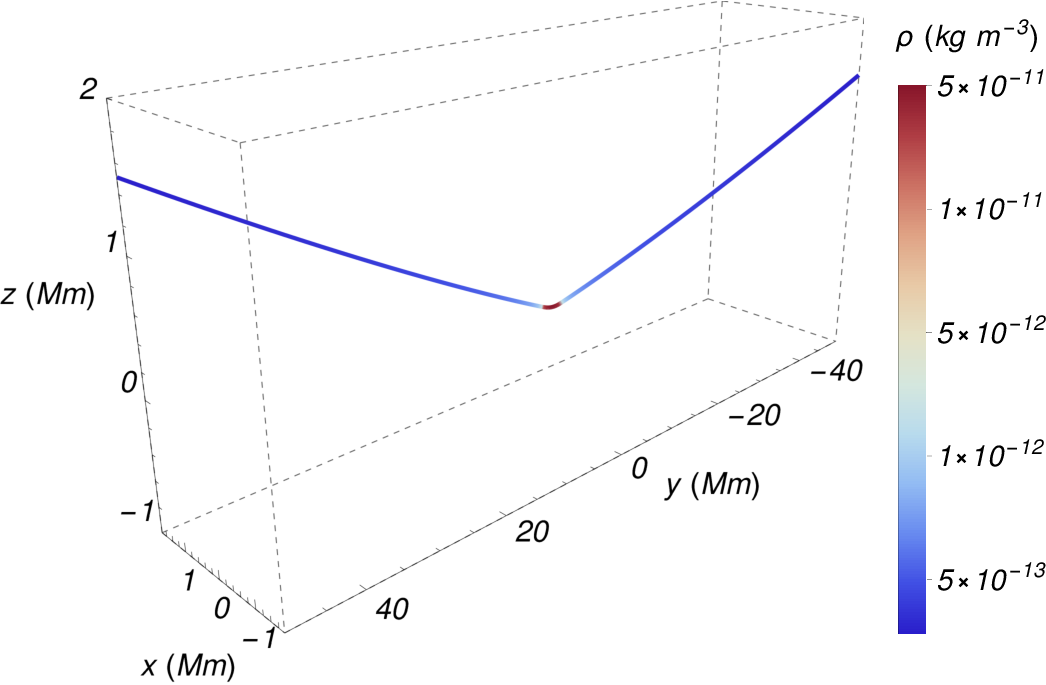}
    \caption{Visualization in 3D of the magnetic dip obtained for the reference model with $T_{0}=$~8,000~K, $p_{0}=5\times 10^{-3}$~Pa, $B_{0}=10$~G, and $\phi=88^{\circ}$. The color gradient denotes the variation of the density along the magnetic field line. Note that the axes are not to scale.}
    \label{fig:line}
\end{figure}

We start by presenting the results of a reference  model computed using a particular set of parameters, namely $T_{0}=$~8,000~K, $p_{0}=5\times 10^{-3}$~Pa, $B_{0}=10$~G, and $\phi=88^{\circ}$. A visualization of the magnetic field line obtained in this configuration is displayed in Figure~\ref{fig:line}, which shows the formation of a dip where the dense prominence plasma is supported.

\begin{figure*}
    \centering
    \includegraphics[width=\linewidth]{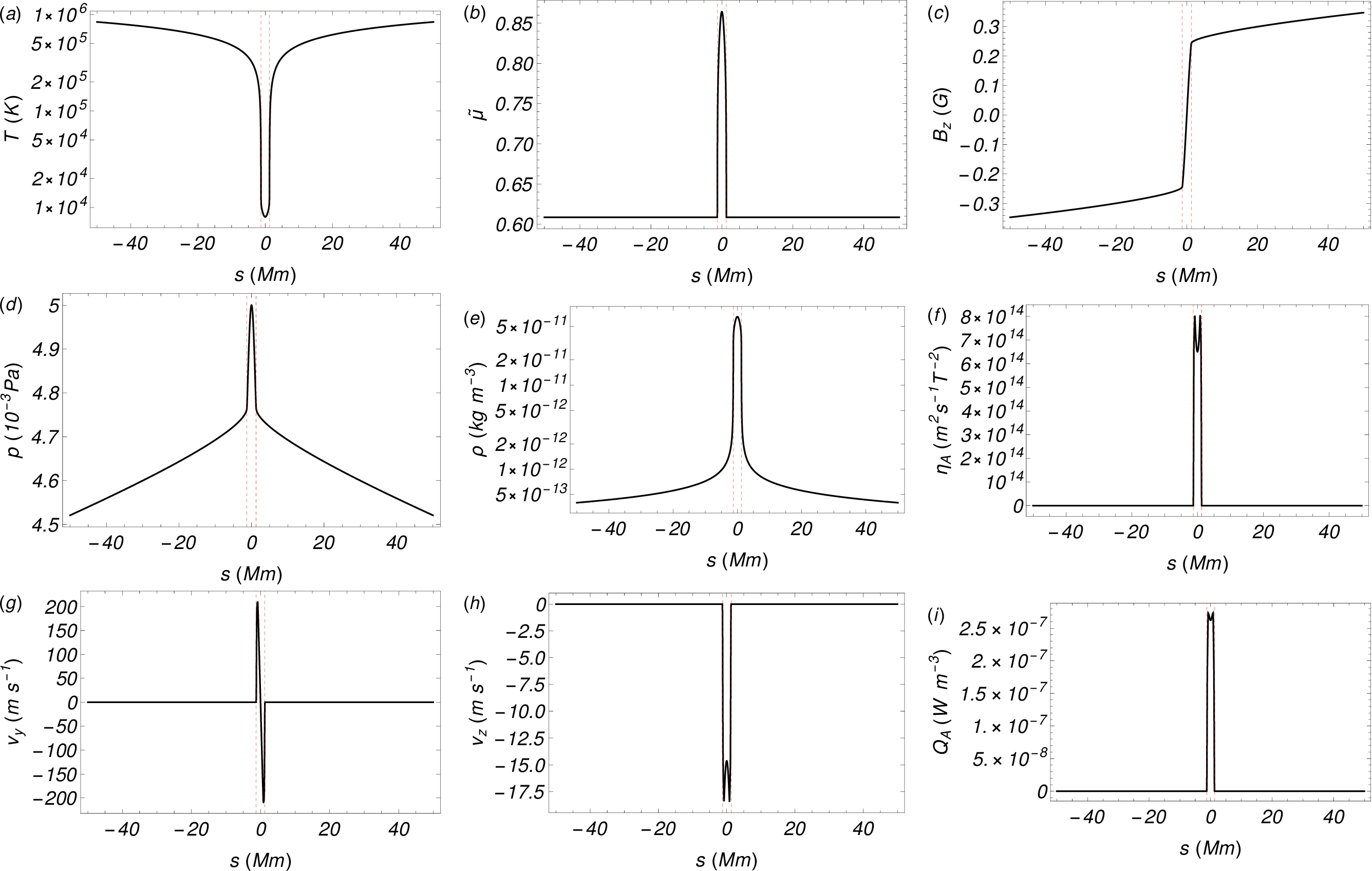}
    \caption{Equilibrium profiles along the magnetic field line for the reference model: a) temperature, b) mean atomic weight, c) $z$-component of the magnetic field, d) pressure, e) density, f) ambipolar diffusion coefficient, g) $y$-component of the velocity, h) $z$-component of the velocity, and i) ambipolar heating rate. Results for $T_{0}=8000$ K, $p_{0}=5\times 10^{-3}$ Pa, $B_{0}=10$ G and $\phi=88^{\circ}$. The vertical red dashed lines mark the location of the cool and dense prominence thread of length $a \approx 2.56$~Mm, being this the distance between the two  lines.}
    \label{fig:magnitudes}
\end{figure*}

Figure~\ref{fig:magnitudes} shows the profiles along the magnetic field line of temperature, density, pressure, mean atomic weight, vertical component of the magnetic field,  ambipolar diffusion coefficient,  $y$- and $z$-components of the velocity, and ambipolar heating rate. The temperature profile (Fig.~\ref{fig:magnitudes}a)  has a shape similar to the profiles presented in \citet{terradas2021thread} and \citet{melis2023heat}, even though these previous works considered very different heating mechanisms: \citet{terradas2021thread} assumed an arbitrary  constant heating and \citet{melis2023heat} considered heating by Alfv\'en waves. The profile can be understood as formed by three different regions: a central cool region with an almost parabolic shape that represents the cold prominence thread, a sharp increase of the temperature by 2 orders of magnitude in a very thin prominence-corona transition region (PCTR), and finally an extended external region where the temperature increase is smoother, reaching typical coronal temperatures of  $\sim 8.5\times 10^{5}$ K. The  length of the cold thread, denoted by $a$, is computed as in  \citet{melis2023heat} by detecting the location of the two PCTR that symmetrically surround the cold region. Due to the considered boundary conditions, the  profile of temperature and those of the other physical quantities are  symmetric or antisymmetric with respect to $s=0$. For this particular model, we find  $a \approx 2.56$~Mm, which is a very small fraction of the total length of the magnetic field line ($L=100$~Mm), but agrees well with  the thread lengths reported in observations \citep[e.g.,][]{lin2011filaments}.

\begin{figure}
    \centering
    \includegraphics[width=0.9\linewidth]{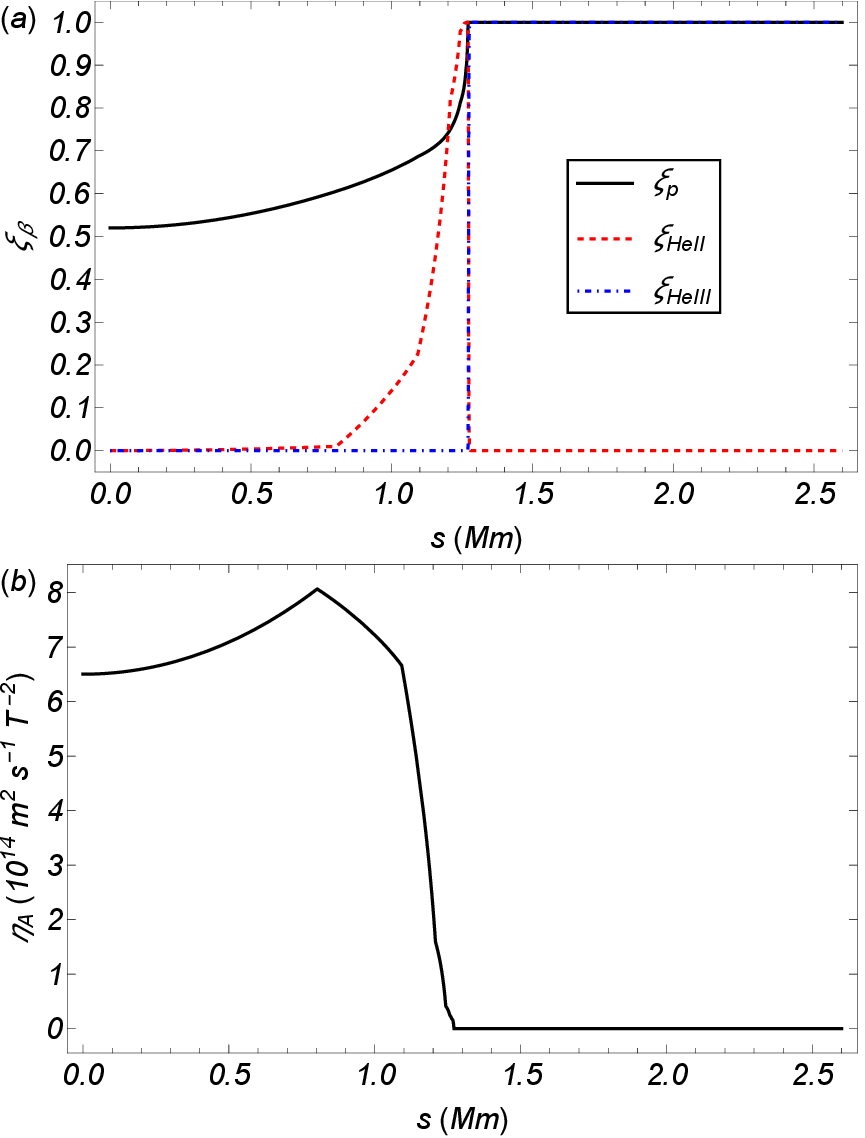}
    \caption{Close-up views of some quantities of the reference model around the cold thread and the PCTR: a) Mass fractions of protons (ionized hydrogen),  singly ionized helium, and doubly ionized helium. b) Ambipolar diffusion coefficient.  Only half the domain with  $s \geq0$ is displayed due to the symmetry of the profiles. The discontinuous derivatives of the ambipolar diffusion coefficient are caused by the first-order interpolation used to implement the hydrogen ionization degree from the tabulated values in \citet{heinzel2015fast}.}
    \label{fig:ion_amb}
\end{figure}

The mean atomic weight, represented in Figure \ref{fig:magnitudes}b, is computed with Eq.~(\ref{eq:weight}). Its behavior is inverse that of the temperature. In the cool central region, $\tilde{\mu}$ reaches its maximum  value of 0.86, approximately, which corresponds to a partially ionized plasma with $\xi_{\rm p} \approx 0.5$, $\xi_{\rm HeII} \approx 10^{-4}$, and $\xi_{\rm HeIII} \approx 0$. This means that, at the thread core, half of hydrogen is ionized, while almost all helium is neutral. In the PCTR, there is a sharp decrease of $\tilde{\mu}$, related with the fast ionization of both hydrogen and helium, and $\tilde{\mu}=1.4/2.3 \approx 0.61$ in the extended coronal region. Such a value corresponds to  a fully ionized hydrogen-helium plasma. A closed-up view of the transition from partial ionization to full ionization is displayed in Figure~\ref{fig:ion_amb}a. The ionization of hydrogen progressively increases with $s$ and reaches its complete ionization at $s\approx1.28$~Mm, within the PCTR. Conversely, the ionization of helium happens in a much thinner region. The fraction of singly ionized helium, $\xi_{\rm HeII}$, peaks within the PCTR and then decreases sharply when doubly ionized helium becomes the dominant helium state. Helium reaches complete ionization at $s \approx 1.29$~Mm, shortly after the location where  complete ionization of hydrogen happens. We recall that the ionization of helium is computed here in a approximate manner with the Saha equation, so that the profiles of $\xi_{\rm HeII}$ and $\xi_{\rm HeIII}$ might be somewhat different in more general approaches.

The vertical component of the magnetic field, computed from Eq.~(\ref{eq:magnetic}), is shown in Figure~\ref{fig:magnitudes}c. We find that $B_z > 0$  for $s>0$ and $B_z < 0$  for $s<0$, so that the profile is antisymmetric with respect to $s=0$, where $B_{z}=0$. This results in the formation of the magnetic dip. The profile has a sharp increase/decrease in the PCTR and a smooth increase/decrease in the outermost region. At the ends of the magnetic field line, where $B_z$ is maximum, we find $B_z \approx \pm 0.35$~G, which is two orders of magnitude below the strength of the horizontal component, $B_{0}$. Therefore, the dip produced by the effect gravity is very shallow. Using the expression for the $z$-coordinate of the magnetic field line (Eq.~(\ref{eq:z})) and setting $z_0=0$ as a reference, we find that $z\approx 1.5$~Mm at $s=L/2$, which confirms that the depth of the dip is much smaller than the length of the magnetic field line.

The pressure profile is calculated from Eq.~(\ref{eq:pressure}) and plotted in Figure~\ref{fig:magnitudes}d. The pressure has its maximum located at the center. Then, it decreases sharply in the PCTR and in a gentler way in the coronal region, although the decrease in the outermost region with respect to the value at the center is only a small percentage. This can be explained by the fact that the variation of $p$ is proportional to $B_z^2$, and $B_z$ itself takes small values, as already discussed. In turn, the density profile is computed from Eq.~(\ref{eq:density}) and represented in Figure~\ref{fig:magnitudes}e. As in the case of the pressure, the density is maximum at the center, with a value of $\rho\approx 5\times 10^{-11}$~kg~m$^{-3}$, and decreases as we move toward the coronal part, where $\rho\sim$10$^{-13}$~kg~m$^{-3}$. The shape of the density profile seems to mimic that of the temperature but with an inverse behavior. The magnetic field line displayed in Figure~\ref{fig:line} has been colored according to the values of the density. The three distinct regions discussed before are clearly shown along the magnetic field line: the dark red region around the dip corresponding to the cold and dense prominence thread, the light blue region denoting the narrow PCTR, and the dark blue representing the evacuated part of the domain matching coronal conditions.

The ambipolar diffusion coefficient, $\eta_{\rm A}$, is plotted in Figure \ref{fig:magnitudes}f. It has a null value in the coronal region, where the plasma is fully ionized. In the cold central region, it has two relative maxima slightly displaced from the center. This shape agrees with the models computed in \citet{melis2023heat}. The maximum value of the coefficient is  $\eta_{\rm A}\approx 10^{14}$ m$^{2}$s$^{-1}$T$^{-2}$, which is an order of magnitude above the values computed in previous works where the plasma was considered to be composed of hydrogen alone \citep[see][]{melis2021heat,melis2023heat}. Here, the effect of helium is included and it has an important contribution to the ambipolar diffusion coefficient. The important role of helium for accurately computing $\eta_{\rm A}$ was discussed by \citet{Zaqarashvili2013} in the context of the chromosphere. A detailed view of $\eta_{\rm A}$ in the cold central region is represented in Figure~\ref{fig:ion_amb}b, where the location of the maximum of $\eta_{\rm A}$ is seen to correlate well with the location where  helium begins to ionize (see the  Figure~\ref{fig:ion_amb}a). The sharp decrease of $\eta_{\rm A}$ within the PCTR, owing to the fact that the plasma gets fully ionized there, is also clearly visible. 

The $y$- and $z$-components of the equilibrium flow velocity are represented in Figures~\ref{fig:magnitudes}g and \ref{fig:magnitudes}h, respectively. Both velocities are caused by ambipolar diffusion and depend on the ambipolar diffusion coefficient, as seen in Eqs.~(\ref{eq:vertical}) and (\ref{eq:drainage}). The $y$-component is antisymmetric with respect to the thread center, being positive for $s<0$ and negative for $s > 0$, taking a maximum value of $v_y\approx\pm200$~m~s$^{-1}$. This means that $v_y$ represents a converging flow that is, essentially, along the magnetic field lines, owing to the large shear angle considered in this model. Despite the presence of this converging flow, the equilibrium density does not change in time. The reason for this result resides in the role of the $z$-component of the velocity. The vertical component of the equilibrium flow takes negative values within the thread, which means that $v_z$ is a drainage flow that causes the fall of some of the prominence material, compensating the convergence of plasma caused by $v_y$. The  value of $v_z$ around the thread center is $v_z\approx -14$~m~s$^{-1}$. This drainage flow we obtain in the model can be compared with that studied in \cite{gilbert2002neutral} and \cite{terradas2015prominence}. In order to compare the present results with those of the previous references, Eq.~(\ref{eq:drainage}) should be rewritten by using the relation between the density and the vertical component of the magnetic field. Since the vertical magnetic field profile is antisymmetric, the drainage velocity at the thread center can be expressed as,
\begin{equation}
    v_{z}= -\mu_{0}g\,\eta_{\rm A}\, \rho, \quad {\rm at} \quad s=0. \label{eq:drain1}
\end{equation}
Moreover, the ambipolar coefficient should be expressed in terms of the collision frequencies. The frequency of collisions between species $\beta$ and $\beta'$ is, 
\begin{equation}
    \nu_{\beta\beta'} = \frac{\alpha_{\beta\beta'}}{\xi_\beta \, \rho},
\end{equation}
while the total collision frequency of species $\beta$ is,
\begin{equation}
    \nu_{\beta} = \sum_{\beta \neq \beta'} \nu_{\beta\beta'}.
\end{equation}
Expressing $\eta_{\rm A}$ from Eq.~(\ref{eq:etaa}) in terms of the collision frequencies, and neglecting collisions between neutral hydrogen and neutral helium, allows us to rewrite Eq.~(\ref{eq:drain1}) as
\begin{equation}
    v_{z} \approx -g \left( \frac{\xi_{\rm H}}{\nu_{\rm H}}+\frac{\xi_{\rm HeI}}{\nu_{\rm HeI}} \right) = \xi_{\rm H}u_{\rm H} + \xi_{\rm HeI}u_{\rm HeI}, \quad {\rm at} \quad s=0, \label{eq:drain2}
\end{equation}
where $u_{\rm H} = -g/\nu_{\rm H}$ and $u_{\rm HeI} = -g/\nu_{\rm HeI}$ denote the drainage velocities for  neutral hydrogen and neutral helium, respectively. Essentially, the total drainage velocity can be calculated as the weighted average of the drainage velocities of the two neutral species, with the weights being their respective abundances.  At the thread center, the collision frequencies are $\nu_{\rm H} \approx 135$~Hz and $\nu_{\rm HeI}\approx 5.75$~Hz, and the drainage velocities for each species are $u_{\rm H} \approx -2$~m~s$^{-1}$ and $u_{\rm HeI} \approx -47.6$~m~s$^{-1}$. The computed  hydrogen drainage velocity matches almost exactly the value of $-2.3$~m~s$^{-1}$ given in \cite{terradas2015prominence}. The computed drainage velocities are also of the same order of magnitude as the results of \cite{gilbert2002neutral}, who obtained drainage velocities of $-3.7$~m~s$^{-1}$ for neutral hydrogen and $-81.1$~m~s$^{-1}$ for neutral helium. Therefore, the model captures well the drainage of neutrals due to the effect of gravity. Observationally, detecting such slow drainage flows is challenging and there is only indirect  evidence of their existence \citep{gilbert2002neutral,gilbert2007}.

Finally, the ambipolar heating rate is plotted in Figure~\ref{fig:magnitudes}i. The largest values of the heating rate are found in the cool prominence region, although not exactly at the center but slightly displaced from it, in a similar fashion as the $\eta_{\rm A}$ profile. This is explained by the fact that $Q_A$ is proportional to $\eta_{\rm A}$, as shown in Eq.~(\ref{eq:heatingrate}). Then the heating rate undergoes a sharp decrease in the PCTR and is absent in the coronal region, where the plasma is fully ionized and $\eta_{\rm A}$ vanishes.

\subsection{Mechanical equilibrium and energy balance}

\begin{figure}
    \centering
    \includegraphics[width=0.9\linewidth]{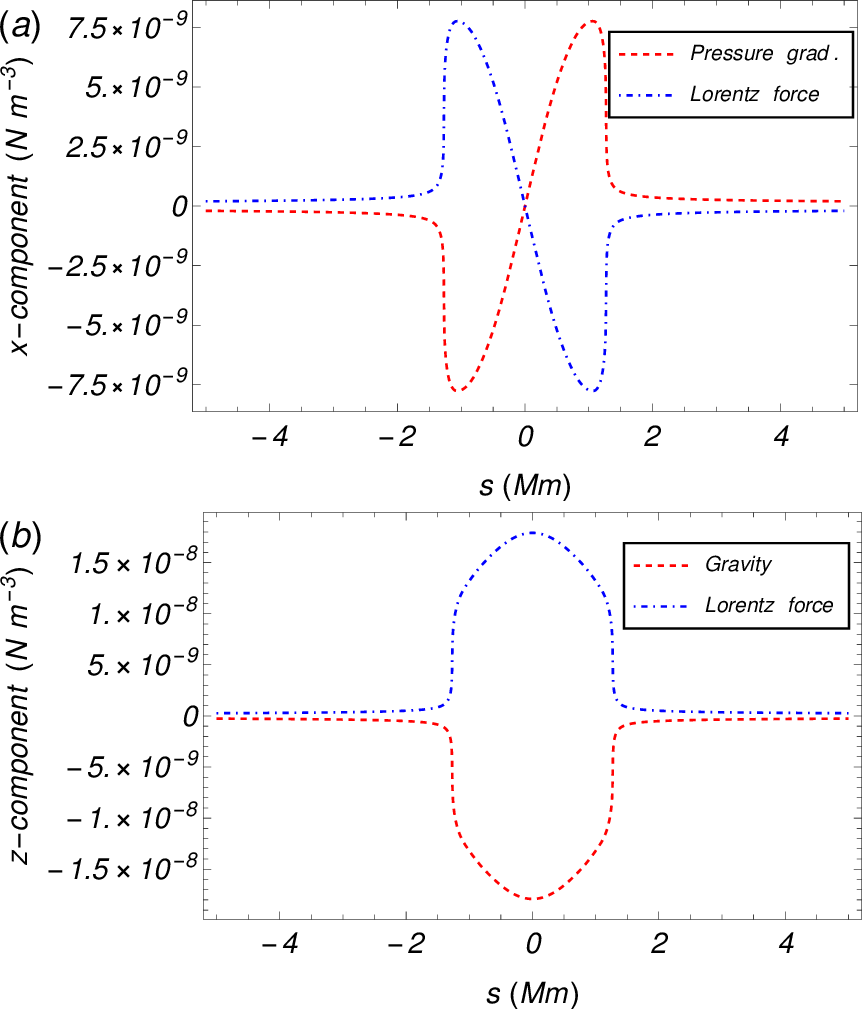}
    \caption{Mechanical equilibrium along the magnetic field line for the reference model:  a) longitudinal and b) vertical components of the forces. Only a region around the cold thread, the PCTR, and the beginning  of the coronal part is displayed.}
    \label{fig:force}
\end{figure}

In this section we explore how the mechanical equilibrium and the energy balance are established. Regarding the  mechanical equilibrium, we computed the forces on the right-hand side of the momentum equation (Eq.~(\ref{eq:momentum}))  in the case of the reference model discussed in the previous section.

Figure~\ref{fig:force} shows the equilibrium of the forces in the $x$-direction (panel a) and the $z$-direction (panel b). This Figure can be compared with Figures~5 and 6 of \citet{terradas2013}, who investigated the balance of forces in a 2D prominence equilibrium. The forces acting in the $x$-direction, i.e., the horizontal direction, are the gas pressure gradient and the $x$-component of the Lorentz force, which is essentially the magnetic pressure gradient. These two forces compensate each other, so that  the total  (gas plus magnetic) pressure remains constant. These pressure gradients are antisymmetric, having their maxima near the PCTR and decreasing sharply in the coronal region. On the other hand, the forces acting in the $z$-direction, i.e., the vertical direction, are gravity and the $z$-component of the Lorentz force, which is essentially the magnetic tension.  As gravity is pointing downward, the magnetic tension needs to provide an upward force to sustain the prominence plasma. Both profiles are symmetric with respect to the thread center and maximum at the center, and decrease sharply in the PCTR and the coronal region. Of course, this behavior is related to the shape and values of the density. Comparing the values of the various forces, we see that the vertical components are up to an order of magnitude larger than  the horizontal components. This result can be attributed to the important role of gravity.

\begin{figure}
    \centering
    \includegraphics[width=0.9\linewidth]{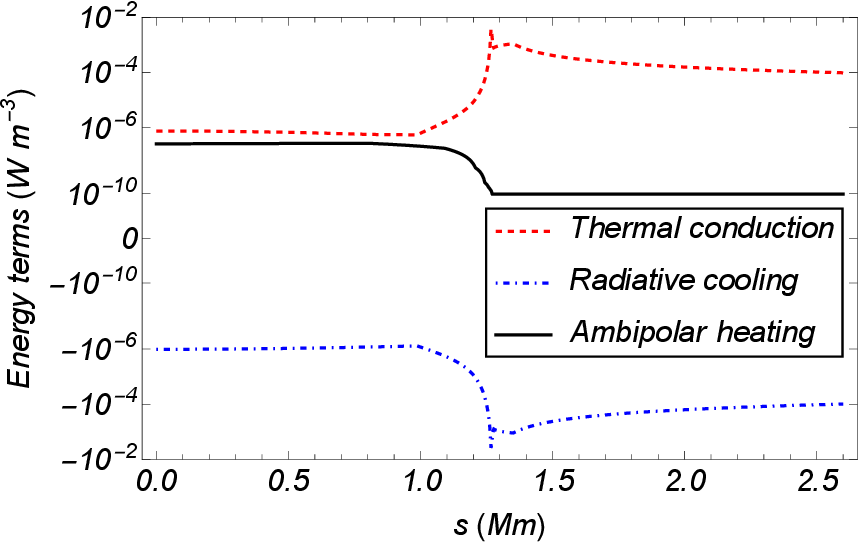}
    \caption{Comparison of the energy terms in the heat-loss function (Eq.~(\ref{eq:mathcalL})) along the magnetic field line for the reference model. Only a region around the cold thread, the PCTR, and the beginning  of the coronal part is displayed. Only half the domain with $s \geq0$ is displayed due to the symmetry of the profiles. In the coronal part, the ambipolar heating is set to a nonzero but  negligible value for numerical stability reasons.}
    \label{fig:energy}
\end{figure}

Concerning the study of the energy balance, we have computed the terms on the right-hand side of Eq.~(\ref{eq:mathcalL}) using, again, the parameters of the reference model. Explicitly, these terms are the ambipolar heating rate, divergence of the heat flux due to thermal conduction, and the energy loss due to radiative cooling. Figure~\ref{fig:energy} compares the relevance of such terms along the magnetic field line in a zone encompassing the cool thread, the PCTR, and the beginning of the coronal region. The sign of each term indicates whether the term acts as an energy sink (negative) term  or  an energy source (positive) term. Obviously, radiative cooling is a sink term that extracts energy from the plasma.   The radiative losses are most efficient within the PCTR, for a temperature of  $T\approx$~57,000~K, and their minimum (in absolute value) takes place in the cold prominence thread. The term due to thermal conduction follows a similar profile as that of radiative losses, specially in the PCTR and the coronal part, but with a positive sign. Finally, the ambipolar heating rate, which was already displayed in Figure~\ref{fig:magnitudes}i, is positive, as expected for a term that provides an energy input. However, this heating term  is  confined within the cool region where the plasma is partially ionized.

Comparing the relevance of each term in the energy balance, we see that  the radiative losses are entirely compensated by the thermal conduction in the PCTR and the coronal part, since the ambipolar heating is  there absent. However, in the cool central region the important energy input provided by ambipolar heating partly compensates for the radiative losses, and the relative weight of the  conduction term in the energy balance consequently decreases. These findings about the relevant role of the heating term in the cool thread are consistent with previous works where either a prescribed uniform heating \citep{terradas2021thread} or a self-consistently computed nonuniform wave heating \citep{melis2023heat} were considered. An important difference with respect to those previous works is that the heating considered here is neither arbitrarily imposed, as in \citet{terradas2021thread}, nor produced by external sources, as in \citet{melis2023heat}; instead, it is generated by the internal diffusion of non–force-free magnetic field of the prominence itself.

\begin{figure}
    \centering
    \includegraphics[width=0.9\linewidth]{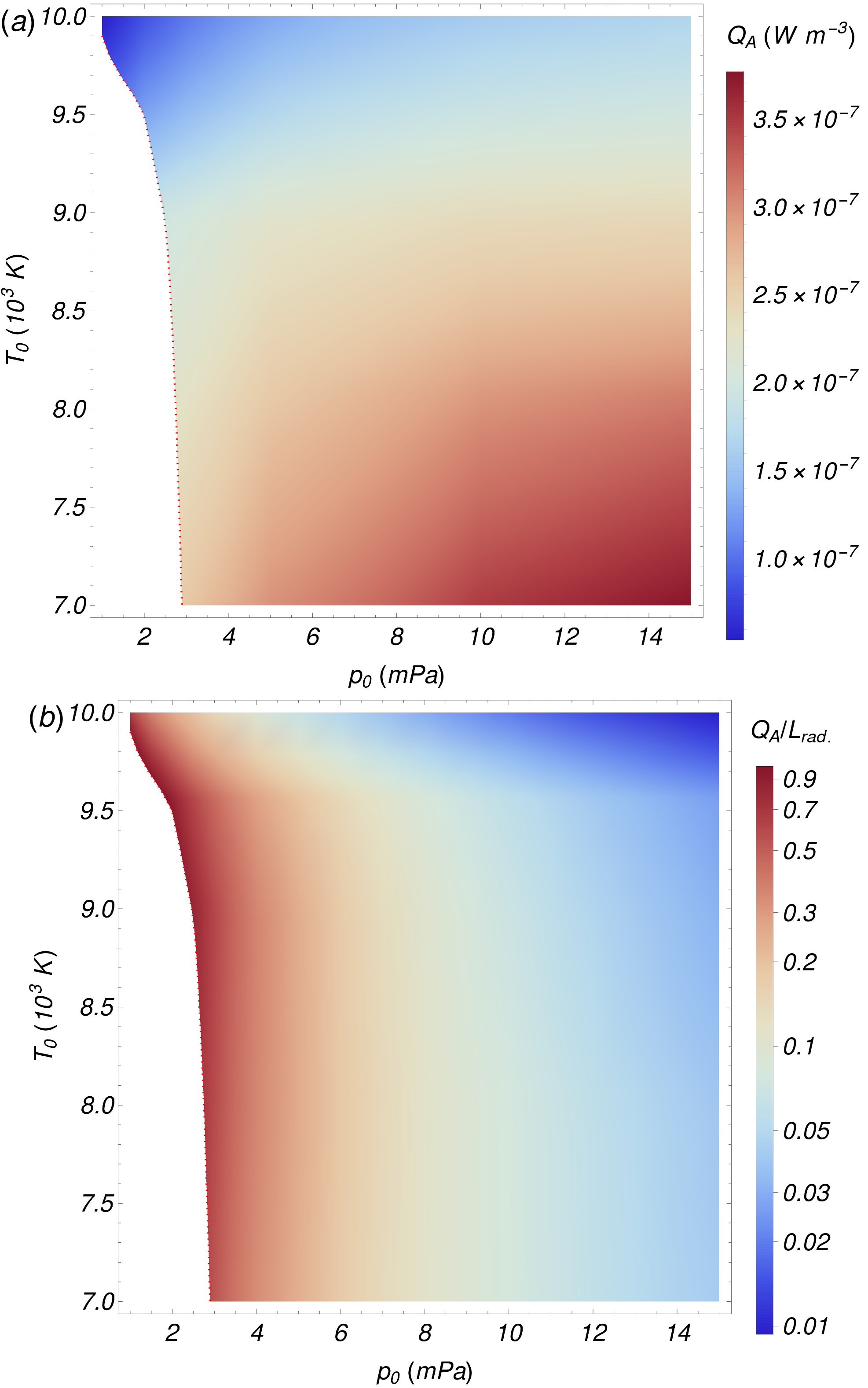}
    \caption{Contour plots of the ambipolar heating at the center (upper panel) and the ratio between heating and cooling at the center (lower panel), as functions of $T_{0}$ and $p_{0}$. The color range in upper panel is in lineal scale whereas the lower panel is in logarithmic scale.}
    \label{fig:hr}
\end{figure}

The central values of the energy terms depend on  $T_{0}$ and $p_{0}$. We have computed additional models in order to study the dependence of the ambipolar heating at the center as a function of these two parameters and its importance compared with radiative losses. These results are displayed in Figure~\ref{fig:hr}  in the form of contour plots. Importantly, for every value of $T_{0}$ there is a minimum value of $p_{0}$ for which an equilibrium is reachable. These critical values are represented with a red dotted line in Figure~\ref{fig:hr} and are explored later in Section~\ref{sec:param}.

Figure~\ref{fig:hr}a  represents the ambipolar heating at the center. The ambipolar heating is largest when the pressure is high and the temperature is low. This can be explained by the indirect relationship between the ambipolar heating and the mass density. According to Eq.~(\ref{eq:gas}), high pressures and low temperatures are associated with large densities. Then, according to Eq.~(\ref{eq:density}), large densities are related with large values of $d B_x/dx$, which, in turn, give rise to large values of $Q_A$ when Eq.~(\ref{eq:heatingrate}) is taken into account. In more plain physical terms, the larger the density, the deeper the magnetic dip needs to be to support the plasma. Consequently, the larger the equilibrium current, and so the larger the heating produced by the dissipation of such current. However, remarkably, the central heating is independent on both the horizontal magnetic field and the shear angle.

On the other hand, the ratio between the central values of the ambipolar heating and the radiative cooling can be seen in Figure~\ref{fig:hr}b for the same computations as before. For low and intermediate central pressures, the ambipolar heating can compensate for a significant fraction of radiative losses. Particularly, for pressures lower than 6~mPa, the ambipolar heating can be as large as 90\% of radiative losses, although for higher pressures the percentage decreases rapidly. For instance, for 10~mPa the ambipolar heating can only compensate for about 10\% of radiative losses. We note that the ratio $Q_{\rm A}/L_{\rm rad.}$ is almost independent of the central temperature, except for large temperatures approaching 10,000~K.

\subsection{Parameter survey}
\label{sec:param}
\begin{figure*}
    \centering
    \includegraphics[width=0.8\linewidth]{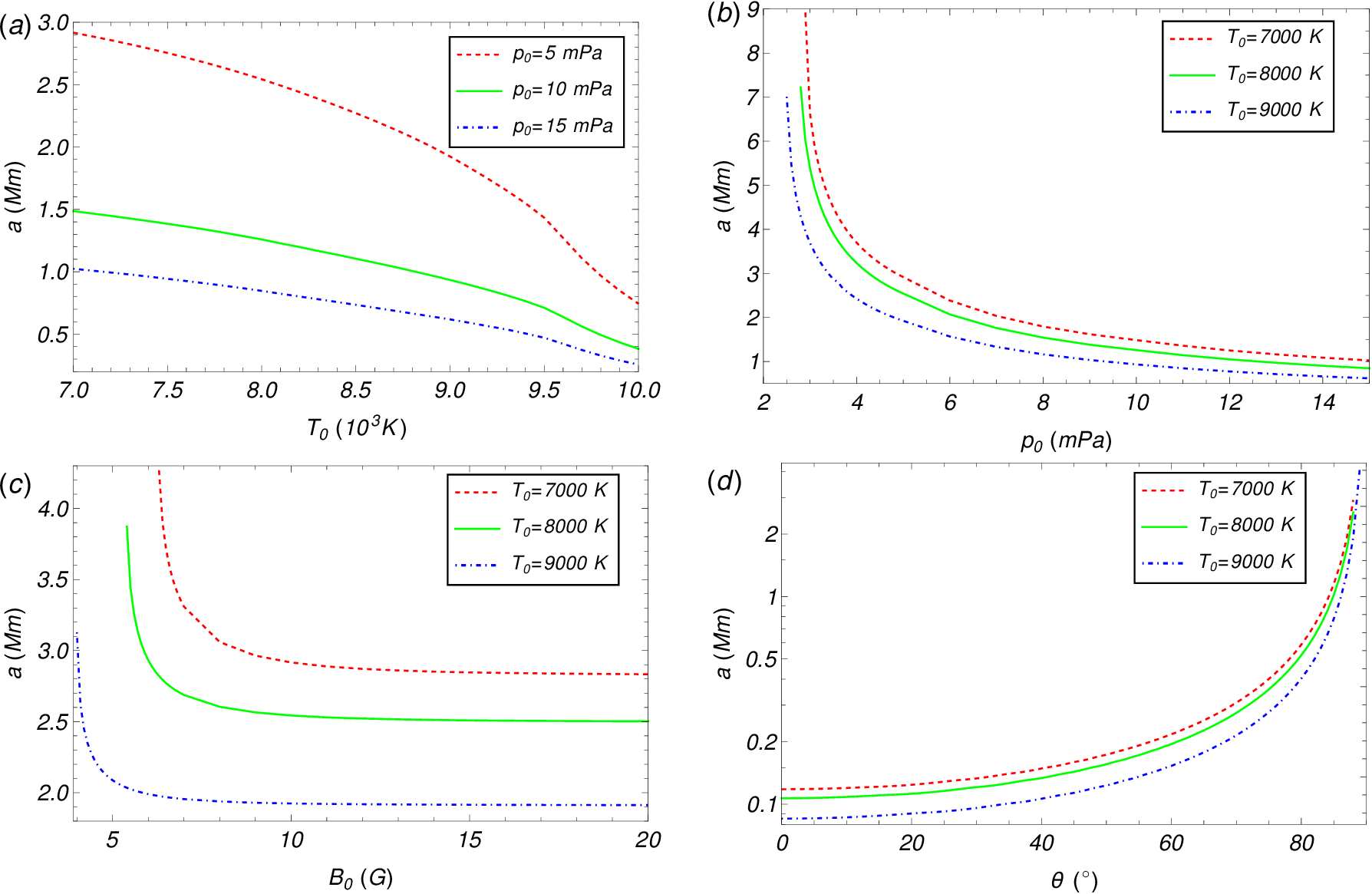}
    \caption{Parameter survey of the thread length, a, in terms of the parameters: $T_{0}$ (upper left panel), $p_{0}$ (upper right), $B_{0}$ (lower left) and $\phi$ (lower right). All the panels are in linear scale with the exception of the lower right panel, which is in logarithmic scale.}
    \label{fig:parameter}
\end{figure*}

Here, we investigate in more detail the effect of changing the free parameters of the model. We recall that these are the central pressure, $p_{0}$, the  horizontal magnetic field strength, $B_{0}$, the shear angle, $\phi$, and the central temperature, $T_{0}$. We focus on analyzing the role of these parameters in the length of the cold thread, $a$. Realistic values of prominence thread lengths reported in observations  are on the order of 2--30 Mm \citep[see, e.g.][]{lin2005,arregui2018}.  The parameter study is performed in the following way. Since there are four free parameters, we calculated the dependence of $a$ with one specific  parameter while keeping the other three parameters unchanged. This parameter survey is summarized in Figure~\ref{fig:parameter}. Unless otherwise stated, the values of the unchanged parameters are same as those of the reference model of Section \ref{sec:typ}.

The effect of $T_0$ is shown in Figure~\ref{fig:parameter}a. Higher central temperatures produce shorter threads, which is related with the increasing efficiency of radiative losses when the temperature increases. This effect  produces steeper temperature gradients in the conduction term and, consequently, shorter threads. This finding is consistent with previous works \citep[see][]{terradas2021thread,melis2023heat}. Figure~\ref{fig:parameter}a contains three curves for three different values of $p_0$, which show that, for the same temperature, the larger the central pressure, the shorter the cold thread. This result is explored in more detail in Figure~\ref{fig:parameter}b, which displays $a$ as a function of $p_0$ for three different values of $T_0$. Hence, Figures~\ref{fig:parameter}a and b are reciprocal to each other. The influence of the central pressure on the thread length results from a complicated interplay between radiative losses and heating, as both ultimately depend on $p_0$ through $\rho$, $B_z$, and the ionization degree. Findings from \citet{terradas2021thread} and \citet{melis2023heat} indicate that threads become shorter when the heating rate decreases in relation with the cooling rate. The result that larger pressures produce shorter threads can be explained by the fact that the heating rate decreases faster than the radiative losses at the core of the thread when $p_0$ increases and $T_0$ remains fixed, as previously shown in Figure~\ref{fig:hr}b. As already mentioned, the value of the central temperature determines a critical value of the central pressure  below which an equilibrium is impossible. The larger $T_0$, the smaller this critical $p_0$. The existence of this critical pressure is related to the largest value of the plasma density that the magnetic field is able to support.

The influence of the strength of the horizontal magnetic field, $B_{0}$,  is shown in Figure \ref{fig:parameter}c. Stronger magnetic fields provide shorter threads, although the length saturates to a constant value once the magnetic field is strong enough. However, for weak magnetic fields the thread length increases asymptotically when $B_0$ decreases, so that there is a critical value of the magnetic field strength for an equilibrium to exist,  in a similar fashion as happens for the central pressure. As before, three different values of the central temperature have been considered, with higher temperatures providing shorter threads, being this consistent with the findings discussed before. Moreover, the central temperature also affects the critical value of the magnetic field strength, so that the critical $B_0$ decreases when $T_0$ increases. As an example, for $T_0=$~7000~K the critical magnetic field strength is $B_0 \approx$~6.3~G, meanwhile for $T_0=$~9000~K this value gets reduced to $B_0 \approx$~4~G. The combined effects of $B_0$ and $p_0$ is analyzed in more detail in Appendix~\ref{sec:beta}.

To end this parameter study, Figure~\ref{fig:parameter}d shows the effect of the magnetic shear angle. The thread length increases for increasing shear angles, in a more significant manner when the shear angle approaches 90$^{\circ}$. For small magnetic shear, the thread length has unrealistically small values of less than 0.2~Mm. In this configuration, large magnetic shear is needed to obtain realistic values of the thread length of the order of 2~Mm or larger. We note that the equilibrium models are computed along the $x$-direction and then  projected in the $s$-direction, i.e., the direction of the magnetic field, according the the prescribed shear. Therefore, the total thread length depends on the shear angle through the relation $a = a_{x}/\cos{\phi}$, where $a_{x}$ denotes the thread length measured along the $x$-direction, which is weakly modified when varying $\phi$.

\section{Discussion and caveats}
\label{sec:conclusions}

In this paper we have explored the role of the heating produced by  ambipolar diffusion in the energy balance of solar prominences, using an idealized  1D model. We considered a modified K-S configuration in order to construct models that satisfy both mechanical and energy equilibria, with the effect of ambipolar heating included in a consistent manner. We used a numerical method, based on that of \cite{melis2023heat}, where the balance of forces and the energy balance are solved iteratively until convergence to a self-consistent model was achieved.

The obtained models  have a similar structure as those computed previously in \citet{terradas2021thread} and \cite{melis2023heat}. A cool prominence region was found at the center, followed by a very thin PCTR, and an extended coronal region. The dense prominence plasma is supported against gravity by a shallow magnetic dip. In the cool prominence region, the plasma is partially ionized and ambipolar diffusion produces heating by dissipating currents associated with the non-force-free magnetic field. Additionally, ambipolar diffusion drives stationary flows associated with the gravitational drainage of neutrals.

A parameter survey was done to explore the variation of the length of the cold region. Cold regions with lengths compatible with those observed are obtained when realistic temperatures, pressures, and magnetic field strengths are considered, although the shear angle has an important effect. Only large values of the shear angle provide realistic cold thread lengths.  In turn,  there are critical values of $p_0$, $T_0$, and $B_{0}$ for which computing an equilibrium configuration is no longer possible, and the length of the cold region increases asymptotically when those critical values are  approached. Appendix~\ref{sec:beta} expands and complements the results of the parameter survey.

The force  balance has been analyzed, showing that, essentially, magnetic tension compensates gravity in  the vertical direction, whereas magnetic pressure balances gas pressure in the horizontal direction. Regarding the energy balance, it has been demonstrated that the heating produced by  ambipolar diffusion can partly compensate for the cooling  by radiation. The ambipolar heating can balance almost entirely the radiative losses for very hot and tenuous threads. Considering more realistic values of temperature, pressure, and density, it is more conservative to state that ambipolar heating can compensate for, at least, 10\% of radiative losses. We therefore conclude that the heating provided by the ambipolar diffusion of the prominence own magnetic field can play a non-negligible role in the energy balance of prominences.

The models explored here have some limitations that can be improved in the future. We used a 1D configuration, which obviate  complexities that may appear in multi-dimensional models. An extension could be made by studying prominence threads in cylindrical geometry, which can include more ingredients such as the perpendicular component of  thermal conduction. In this context, the study of thread widths in multi-dimensional models would be a natural extension of this work. Another limitation is that the models are stationary.  However,  prominences are dynamic objects that can suffer changes in relatively short periods of time. The models constructed here apply only to the global, large-scale quasi-equilibrium of prominences and ignore the rich temporal evolution that is especially important at small scales. Another aspect we have not addressed is the stability of the constructed models. In this regard, the stability analysis of \citet{Bakhareva1992}, including partial ionization effects, could be a starting point for future research. In this direction, studying the eigenmodes of the equilibrium models can help determining their stability, as in \cite{schmitt1995}.

The present results can be qualitatively compared with those of recent time-dependent studies of prominence and coronal rain formation, allowing potentially important effects not included here to be identified. For instance, \cite{jercic2025} studied prominence formation in a 1D hydrodynamic setup using the two-fluid theory. Although their computed structures show similarities with our models in terms of temperature, density, and pressure profiles, the formation and evolution of shocks play a prominent role in their simulations. Such dynamics cannot be captured by our stationary models. Additionally, the threads obtained in the simulations continuously accrete mass and increase their lengths beyond the limited lengths obtained from our equilibrium models, and reconciling the results of these different approaches requires further analysis. Similarly, \cite{popescu2025} used the two-fluid theory to study coronal rain formation in a 3D setup without gravity. Their results show that ionization and recombination processes play an important role in their simulations. However, ionization and recombination cannot be described with the single-fluid approach used here.

Although our focus has been put on studying the influence of ambipolar heating and not on radiative losses, a potential limitation of this work is the approximate treatment of the cooling rates. We can roughly compare the radiative cooling rates obtained at the center of the reference model using the approximate cooling function (Eq.~(\ref{eq:radiation})) with those given in \citet{gunar2025radiative}. To this end, we consider their results for a 1D vertical slab illuminated from its sides, and use $T=$~8,000~K and $p=$~0.05~dyn~cm$^{-2}$. To estimate the distance to the  illuminated surface, we take into account that the length of the cold thread is 2.56~Mm, which means that the PCTR is located at a distance of 1.28~Mm from the center. Approximating the location of the PCTR as the location of the slab surface in the model of \citet{gunar2025radiative}, we use $L\approx$~1250~km in their Table~A.5. The tabulated cooling rate is $\sim 2 \times 10^{-5}$~W~m$^{-3}$, which is about a factor of 10 larger than that computed from Eq.~(\ref{eq:radiation}). Therefore,  using the net radiative cooling rates of  \citet{gunar2025radiative} may introduce some differences in the properties of the models discussed here. We plan to explore this in detail in a forthcoming work.

\begin{acknowledgements}
      This publication is part of the R+D+i project PID2023-147708NB-I00, funded by MCIN/AEI/10.13039/501100011033 and by FEDER, EU. LM is supported by the predoctoral fellowship FPI\_002\_2022  funded by CAIB. We thank Prof. J. L. Ballester for reading a draft of this paper and for providing useful comments. We thank the anonymous referee for useful comments.
\end{acknowledgements}

\bibpunct{(}{)}{;}{a}{}{,} 

\bibliographystyle{aa} 
\bibliography{refs} 

@ARTICLE{melis2021heat,
       author = {{Melis}, Lloren{\c{c}} and {Soler}, Roberto and {Ballester}, Jos{\'e} Luis},
        title = "{Alfv{\'e}n wave heating in partially ionized thin threads of solar prominences}",
      journal = {\aap},
         year = 2021,
        month = jun,
       volume = {650},
          eid = {A45},
        pages = {A45},
          doi = {10.1051/0004-6361/202140523},
archivePrefix = {arXiv},
       eprint = {2103.16599},
 primaryClass = {astro-ph.SR},
       adsurl = {https://ui.adsabs.harvard.edu/abs/2021A&A...650A..45M}
}

@ARTICLE{melis2023heat,
       author = {{Melis}, Lloren{\c{c}} and {Soler}, Roberto and {Terradas}, Jaume},
        title = "{Self-consistent equilibrium models of prominence thin threads heated by Alfv{\'e}n waves propagating from the photosphere}",
      journal = {\aap},
         year = 2023,
        month = aug,
       volume = {676},
          eid = {A25},
        pages = {A25},
          doi = {10.1051/0004-6361/202346459},
archivePrefix = {arXiv},
       eprint = {2306.13434},
 primaryClass = {astro-ph.SR},
       adsurl = {https://ui.adsabs.harvard.edu/abs/2023A&A...676A..25M}
}

@ARTICLE{terradas2021thread,
       author = {{Terradas}, J. and {Luna}, M. and {Soler}, R. and {Oliver}, R. and {Carbonell}, M. and {Ballester}, J.~L.},
        title = "{One-dimensional prominence threads. I. Equilibrium models}",
      journal = {\aap},
         year = 2021,
        month = sep,
       volume = {653},
          eid = {A95},
        pages = {A95},
          doi = {10.1051/0004-6361/202039905},
archivePrefix = {arXiv},
       eprint = {2106.06327},
 primaryClass = {astro-ph.SR},
       adsurl = {https://ui.adsabs.harvard.edu/abs/2021A&A...653A..95T}
}

@ARTICLE{heinzel2015fast,
       author = {{Heinzel}, P. and {Gun{\'a}r}, S. and {Anzer}, U.},
        title = "{Fast approximate radiative transfer method for visualizing the fine structure of prominences in the hydrogen H{\ensuremath{\alpha}} line}",
      journal = {\aap},
         year = 2015,
        month = jul,
       volume = {579},
          eid = {A16},
        pages = {A16},
          doi = {10.1051/0004-6361/201525716},
       adsurl = {https://ui.adsabs.harvard.edu/abs/2015A&A...579A..16H}
}

@ARTICLE{hermans2021cooling,
       author = {{Hermans}, J. and {Keppens}, R.},
        title = "{Effect of optically thin cooling curves on condensation formation: Case study using thermal instability}",
      journal = {\aap},
         year = 2021,
        month = nov,
       volume = {655},
          eid = {A36},
        pages = {A36},
          doi = {10.1051/0004-6361/202140665},
archivePrefix = {arXiv},
       eprint = {2107.07569},
 primaryClass = {astro-ph.SR},
       adsurl = {https://ui.adsabs.harvard.edu/abs/2021A&A...655A..36H}
}

@ARTICLE{schure2009spex,
       author = {{Schure}, K.~M. and {Kosenko}, D. and {Kaastra}, J.~S. and {Keppens}, R. and {Vink}, J.},
        title = "{A new radiative cooling curve based on an up-to-date plasma emission code}",
      journal = {\aap},
         year = 2009,
        month = dec,
       volume = {508},
       number = {2},
        pages = {751-757},
          doi = {10.1051/0004-6361/200912495},
archivePrefix = {arXiv},
       eprint = {0909.5204},
 primaryClass = {astro-ph.GA},
       adsurl = {https://ui.adsabs.harvard.edu/abs/2009A&A...508..751S}
}

@PROCEEDINGS{vial2015prominences,
        title = "{Solar Prominences}",
    booktitle = {Solar Prominences},
         year = 2015,
       editor = {{Vial}, Jean-Claude and {Engvold}, Oddbj{\o}rn},
       series = {Astrophysics and Space Science Library},
       volume = {415},
        month = jan,
          doi = {10.1007/978-3-319-10416-4},
       adsurl = {https://ui.adsabs.harvard.edu/abs/2015ASSL..415.....V}
}

@INPROCEEDINGS{gilbert2015balance,
       author = {{Gilbert}, Holly},
        title = "{Energy Balance}",
    booktitle = {Solar Prominences},
         year = 2015,
       editor = {{Vial}, Jean-Claude and {Engvold}, Oddbj{\o}rn},
       series = {Astrophysics and Space Science Library},
       volume = {415},
        month = jan,
        pages = {157},
          doi = {10.1007/978-3-319-10416-4_7},
       adsurl = {https://ui.adsabs.harvard.edu/abs/2015ASSL..415..157G}
}

@ARTICLE{parenti2014prominences,
       author = {{Parenti}, Susanna},
        title = "{Solar Prominences: Observations}",
      journal = {Living Reviews in Solar Physics},
         year = 2014,
        month = dec,
       volume = {11},
       number = {1},
          eid = {1},
        pages = {1},
          doi = {10.12942/lrsp-2014-1},
       adsurl = {https://ui.adsabs.harvard.edu/abs/2014LRSP...11....1P}
}

@ARTICLE{anzer1999balance,
       author = {{Anzer}, U. and {Heinzel}, P.},
        title = "{The energy balance in solar prominences}",
      journal = {\aap},
         year = 1999,
        month = sep,
       volume = {349},
        pages = {974-984},
       adsurl = {https://ui.adsabs.harvard.edu/abs/1999A&A...349..974A}
}

@ARTICLE{heasley1976strucutre,
       author = {{Heasley}, J.~N. and {Mihalas}, D.},
        title = "{Structure and spectrum of quiescent prominences: energy balance and hydrogen spectrum.}",
      journal = {\apj},
         year = 1976,
        month = apr,
       volume = {205},
        pages = {273-285},
          doi = {10.1086/154273},
       adsurl = {https://ui.adsabs.harvard.edu/abs/1976ApJ...205..273H}
}

@ARTICLE{heinzel2010prominences,
       author = {{Heinzel}, P. and {Anzer}, U. and {Gun{\'a}r}, S.},
        title = "{Solar quiescent prominences. Filamentary structure and energetics}",
      journal = {\memsai},
         year = 2010,
        month = jan,
       volume = {81},
        pages = {654},
       adsurl = {https://ui.adsabs.harvard.edu/abs/2010MmSAI..81..654H}
}

@ARTICLE{heinzel2012radiative,
       author = {{Heinzel}, P. and {Anzer}, U.},
        title = "{Radiative equilibrium in solar prominences reconsidered}",
      journal = {\aap},
         year = 2012,
        month = mar,
       volume = {539},
          eid = {A49},
        pages = {A49},
          doi = {10.1051/0004-6361/200913537},
       adsurl = {https://ui.adsabs.harvard.edu/abs/2012A&A...539A..49H}
}

@ARTICLE{lin2011filaments,
       author = {{Lin}, Yong},
        title = "{Filament Thread-like Structures and Their Small-amplitude Oscillations}",
      journal = {\ssr},
         year = 2011,
        month = jul,
       volume = {158},
       number = {2-4},
        pages = {237-266},
          doi = {10.1007/s11214-010-9672-9},
       adsurl = {https://ui.adsabs.harvard.edu/abs/2011SSRv..158..237L}
}

@INPROCEEDINGS{martin2015prominences,
       author = {{Martin}, Sara F.},
        title = "{The Magnetic Field Structure of Prominences from Direct and Indirect Observations}",
    booktitle = {Solar Prominences},
         year = 2015,
       editor = {{Vial}, Jean-Claude and {Engvold}, Oddbj{\o}rn},
       series = {Astrophysics and Space Science Library},
       volume = {415},
        month = jan,
        pages = {205},
          doi = {10.1007/978-3-319-10416-4_9},
       adsurl = {https://ui.adsabs.harvard.edu/abs/2015ASSL..415..205M}
}

@ARTICLE{degenhardt1993flux,
       author = {{Degenhardt}, U. and {Deinzer}, W.},
        title = "{A flux tube-model for solar prominences}",
      journal = {\aap},
         year = 1993,
        month = oct,
       volume = {278},
       number = {1},
        pages = {288-292},
       adsurl = {https://ui.adsabs.harvard.edu/abs/1993A&A...278..288D}
}

@ARTICLE{mandrini2000currents,
       author = {{Mandrini}, C.~H. and {D{\'e}moulin}, P. and {Klimchuk}, J.~A.},
        title = "{Magnetic Field and Plasma Scaling Laws: Their Implications for Coronal Heating Models}",
      journal = {\apj},
         year = 2000,
        month = feb,
       volume = {530},
       number = {2},
        pages = {999-1015},
          doi = {10.1086/308398},
       adsurl = {https://ui.adsabs.harvard.edu/abs/2000ApJ...530..999M}
}

@ARTICLE{soler2016heat,
       author = {{Soler}, Roberto and {Terradas}, Jaume and {Oliver}, Ramon and {Ballester}, Jose Luis},
        title = "{The role of Alfv{\'e}n wave heating in solar prominences}",
      journal = {\aap},
         year = 2016,
        month = jul,
       volume = {592},
          eid = {A28},
        pages = {A28},
          doi = {10.1051/0004-6361/201628722},
archivePrefix = {arXiv},
       eprint = {1605.07048},
 primaryClass = {astro-ph.SR},
       adsurl = {https://ui.adsabs.harvard.edu/abs/2016A&A...592A..28S}
}

@ARTICLE{brughmans2022rope,
       author = {{Brughmans}, N. and {Jenkins}, J.~M. and {Keppens}, R.},
        title = "{The influence of flux rope heating models on solar prominence formation}",
      journal = {\aap},
         year = 2022,
        month = dec,
       volume = {668},
          eid = {A47},
        pages = {A47},
          doi = {10.1051/0004-6361/202244071},
archivePrefix = {arXiv},
       eprint = {2210.13195},
 primaryClass = {astro-ph.SR},
       adsurl = {https://ui.adsabs.harvard.edu/abs/2022A&A...668A..47B}
}

@ARTICLE{kippenhahn1957filament,
       author = {{Kippenhahn}, R. and {Schl{\"u}ter}, A.},
        title = "{Eine Theorie der solaren Filamente. Mit 7 Textabbildungen}",
      journal = {\zap},
         year = 1957,
        month = jan,
       volume = {43},
        pages = {36},
       adsurl = {https://ui.adsabs.harvard.edu/abs/1957ZA.....43...36K}
}

@ARTICLE{poland1971energy,
       author = {{Poland}, A. and {Anzer}, U.},
        title = "{Energy Balance in Cool Quiescent Prominences}",
      journal = {\solphys},
         year = 1971,
        month = sep,
       volume = {19},
       number = {2},
        pages = {401-413},
          doi = {10.1007/BF00146067},
       adsurl = {https://ui.adsabs.harvard.edu/abs/1971SoPh...19..401P}
}

@ARTICLE{milne1979prominences,
       author = {{Milne}, A.~M. and {Priest}, E.~R. and {Roberts}, B.},
        title = "{A model for quiescent solar prominences.}",
      journal = {\apj},
         year = 1979,
        month = aug,
       volume = {232},
        pages = {304-317},
          doi = {10.1086/157290},
       adsurl = {https://ui.adsabs.harvard.edu/abs/1979ApJ...232..304M}
}

@ARTICLE{ballester1987model,
       author = {{Ballester}, J.~L. and {Priest}, E.~R.},
        title = "{A Two-Dimensional Model for a Solar Prominence}",
      journal = {\solphys},
         year = 1987,
        month = sep,
       volume = {109},
       number = {2},
        pages = {335-349},
          doi = {10.1007/BF00160656},
       adsurl = {https://ui.adsabs.harvard.edu/abs/1987SoPh..109..335B}
}

@ARTICLE{heinzel2001prominence,
       author = {{Heinzel}, P. and {Anzer}, U.},
        title = "{Prominence fine structures in a magnetic equilibrium: Two-dimensional models with multilevel radiative transfer}",
      journal = {\aap},
         year = 2001,
        month = sep,
       volume = {375},
        pages = {1082-1090},
          doi = {10.1051/0004-6361:20010926},
       adsurl = {https://ui.adsabs.harvard.edu/abs/2001A&A...375.1082H}
}

@INPROCEEDINGS{vigh2018model,
       author = {{Vigh}, Carlos D. and {Gonzalez}, Rafael and {Rial}, Diego},
        title = "{Exploring the Kipenhahn-Schl{\"u}ter model}",
    booktitle = {Journal of Physics Conference Series},
         year = 2018,
       series = {Journal of Physics Conference Series},
       volume = {1031},
        month = may,
    publisher = {IOP},
          eid = {012012},
        pages = {012012},
          doi = {10.1088/1742-6596/1031/1/012012},
       adsurl = {https://ui.adsabs.harvard.edu/abs/2018JPhCS1031a2012V}
}

@ARTICLE{hillier2010magnetic,
       author = {{Hillier}, Andrew and {Shibata}, Kazunari and {Isobe}, Hiroaki},
        title = "{Evolution of the Kippenhahn-Schl{\"u}ter Prominence Model Magnetic Field under Cowling Resistivity}",
      journal = {\pasj},
         year = 2010,
        month = oct,
       volume = {62},
        pages = {1231-1237},
          doi = {10.1093/pasj/62.5.1231},
archivePrefix = {arXiv},
       eprint = {1007.1909},
 primaryClass = {astro-ph.SR},
       adsurl = {https://ui.adsabs.harvard.edu/abs/2010PASJ...62.1231H}
}

@ARTICLE{oliver1992waves,
       author = {{Oliver}, R. and {Ballester}, J.~L. and {Hood}, A.~W. and {Priest}, E.~R.},
        title = "{Magnetohydrodynamic Waves in a Solar Prominence}",
      journal = {\apj},
         year = 1992,
        month = nov,
       volume = {400},
        pages = {369},
          doi = {10.1086/172003},
       adsurl = {https://ui.adsabs.harvard.edu/abs/1992ApJ...400..369O}
}

@ARTICLE{oliver1993oscillations,
       author = {{Oliver}, R. and {Ballester}, J.~L. and {Hood}, A.~W. and {Priest}, E.~R.},
        title = "{Oscillations of a Quiescent Solar Prominence Embedded in a Hot Corona}",
      journal = {\apj},
         year = 1993,
        month = jun,
       volume = {409},
        pages = {809},
          doi = {10.1086/172711},
       adsurl = {https://ui.adsabs.harvard.edu/abs/1993ApJ...409..809O}
}

@ARTICLE{low2005structure,
       author = {{Low}, B.~C. and {Petrie}, G.~J.~D.},
        title = "{The Internal Structures and Dynamics of Solar Quiescent Prominences}",
      journal = {\apj},
         year = 2005,
        month = jun,
       volume = {626},
       number = {1},
        pages = {551-562},
          doi = {10.1086/430046},
       adsurl = {https://ui.adsabs.harvard.edu/abs/2005ApJ...626..551L}
}

@ARTICLE{ballester2018partial,
       author = {{Ballester}, Jos{\'e} Luis and {Alexeev}, Igor and {Collados}, Manuel and {Downes}, Turlough and {Pfaff}, Robert F. and {Gilbert}, Holly and {Khodachenko}, Maxim and {Khomenko}, Elena and {Shaikhislamov}, Ildar F. and {Soler}, Roberto and {V{\'a}zquez-Semadeni}, Enrique and {Zaqarashvili}, Teimuraz},
        title = "{Partially Ionized Plasmas in Astrophysics}",
      journal = {\ssr},
         year = 2018,
        month = mar,
       volume = {214},
       number = {2},
          eid = {58},
        pages = {58},
          doi = {10.1007/s11214-018-0485-6},
archivePrefix = {arXiv},
       eprint = {1707.07975},
 primaryClass = {astro-ph.SR},
       adsurl = {https://ui.adsabs.harvard.edu/abs/2018SSRv..214...58B}
}

@ARTICLE{khomenko2012heating,
       author = {{Khomenko}, E. and {Collados}, M.},
        title = "{Heating of the Magnetized Solar Chromosphere by Partial Ionization Effects}",
      journal = {\apj},
         year = 2012,
        month = mar,
       volume = {747},
       number = {2},
          eid = {87},
        pages = {87},
          doi = {10.1088/0004-637X/747/2/87},
archivePrefix = {arXiv},
       eprint = {1112.3374},
 primaryClass = {astro-ph.SR},
       adsurl = {https://ui.adsabs.harvard.edu/abs/2012ApJ...747...87K}
}

@ARTICLE{gunar2025radiative,
       author = {{Gun{\'a}r}, S. and {Heinzel}, P. and {Anzer}, U.},
        title = "{Net radiative cooling rates and partial ionization in cool coronal condensations}",
      journal = {\aap},
         year = 2025,
        month = jul,
       volume = {699},
          eid = {A89},
        pages = {A89},
          doi = {10.1051/0004-6361/202553909},
       adsurl = {https://ui.adsabs.harvard.edu/abs/2025A&A...699A..89G}
}

@ARTICLE{terradas2015prominence,
       author = {{Terradas}, J. and {Soler}, R. and {Oliver}, R. and {Ballester}, J.~L.},
        title = "{On the Support of Neutrals Against Gravity in Solar Prominences}",
      journal = {\apjl},
         year = 2015,
        month = apr,
       volume = {802},
       number = {2},
          eid = {L28},
        pages = {L28},
          doi = {10.1088/2041-8205/802/2/L28},
archivePrefix = {arXiv},
       eprint = {1503.05354},
 primaryClass = {astro-ph.SR},
       adsurl = {https://ui.adsabs.harvard.edu/abs/2015ApJ...802L..28T}
}

@ARTICLE{gilbert2002neutral,
       author = {{Gilbert}, Holly R. and {Hansteen}, Viggo H. and {Holzer}, Thomas E.},
        title = "{Neutral Atom Diffusion in a Partially Ionized Prominence Plasma}",
      journal = {\apj},
         year = 2002,
        month = sep,
       volume = {577},
       number = {1},
        pages = {464-474},
          doi = {10.1086/342165},
       adsurl = {https://ui.adsabs.harvard.edu/abs/2002ApJ...577..464G}
}

@ARTICLE{Khomenko2014,
       author = {{Khomenko}, E. and {Collados}, M. and {D{\'\i}az}, A. and {Vitas}, N.},
        title = "{Fluid description of multi-component solar partially ionized plasma}",
      journal = {Physics of Plasmas},
         year = 2014,
        month = sep,
       volume = {21},
       number = {9},
          eid = {092901},
        pages = {092901},
          doi = {10.1063/1.4894106},
archivePrefix = {arXiv},
       eprint = {1408.1871},
 primaryClass = {astro-ph.SR},
       adsurl = {https://ui.adsabs.harvard.edu/abs/2014PhPl...21i2901K}
}

@ARTICLE{Zaqarashvili2013,
       author = {{Zaqarashvili}, T.~V. and {Khodachenko}, M.~L. and {Soler}, R.},
        title = "{Torsional Alfv{\'e}n waves in partially ionized solar plasma: effects of neutral helium and stratification}",
      journal = {\aap},
         year = 2013,
        month = jan,
       volume = {549},
          eid = {A113},
        pages = {A113},
          doi = {10.1051/0004-6361/201220272},
archivePrefix = {arXiv},
       eprint = {1211.1348},
 primaryClass = {astro-ph.SR},
       adsurl = {https://ui.adsabs.harvard.edu/abs/2013A&A...549A.113Z}
}

@ARTICLE{soler2015,
       author = {{Soler}, Roberto and {Carbonell}, Marc and {Ballester}, Jose Luis},
        title = "{On the Spatial Scales of Wave Heating in the Solar Chromosphere}",
      journal = {\apj},
         year = 2015,
        month = sep,
       volume = {810},
       number = {2},
          eid = {146},
        pages = {146},
          doi = {10.1088/0004-637X/810/2/146},
archivePrefix = {arXiv},
       eprint = {1508.01497},
 primaryClass = {astro-ph.SR},
       adsurl = {https://ui.adsabs.harvard.edu/abs/2015ApJ...810..146S}
}

@ARTICLE{Braginskii1965,
       author = {{Braginskii}, S.~I.},
        title = "{Transport Processes in a Plasma}",
      journal = {Reviews of Plasma Physics},
         year = 1965,
        month = jan,
       volume = {1},
        pages = {205},
       adsurl = {https://ui.adsabs.harvard.edu/abs/1965RvPP....1..205B}
}

@BOOK{Spitzer1962,
       author = {{Spitzer}, L.},
        title = "{Physics of Fully Ionized Gases}",
         year = 1962,
       adsurl = {https://ui.adsabs.harvard.edu/abs/1962pfig.book.....S}
}

@BOOK{Chapman1970,
       author = {{Chapman}, Sydeny and {Cowling}, T.~G.},
        title = "{The mathematical theory of non-uniform gases. an account of the kinetic theory of viscosity, thermal conduction and diffusion in gases}",
         year = 1970,
       adsurl = {https://ui.adsabs.harvard.edu/abs/1970mtnu.book.....C}
}

@ARTICLE{Draine1986,
       author = {{Draine}, B.~T.},
        title = "{Multicomponent, reacting MHD flows}",
      journal = {\mnras},
         year = 1986,
        month = may,
       volume = {220},
        pages = {133-148},
          doi = {10.1093/mnras/220.1.133},
       adsurl = {https://ui.adsabs.harvard.edu/abs/1986MNRAS.220..133D}
}

@ARTICLE{jercic2025,
       author = {{Jer{\v{c}}i{\'c}}, V. and {Popescu Braileanu}, B. and {Keppens}, R.},
        title = "{Forming Prominences Accounting for Partial Ionization Effects}",
      journal = {\apj},
         year = 2025,
        month = jun,
       volume = {986},
       number = {2},
          eid = {134},
        pages = {134},
          doi = {10.3847/1538-4357/add6aa},
archivePrefix = {arXiv},
       eprint = {2505.07990},
 primaryClass = {astro-ph.SR},
       adsurl = {https://ui.adsabs.harvard.edu/abs/2025ApJ...986..134J}
}

@ARTICLE{keppens2025,
       author = {{Keppens}, Rony and {Zhou}, Yuhao and {Xia}, Chun},
        title = "{Modeling multiphase plasma in the corona: prominences and rain}",
      journal = {Living Reviews in Solar Physics},
         year = 2025,
        month = dec,
       volume = {22},
       number = {1},
          eid = {4},
        pages = {4},
          doi = {10.1007/s41116-025-00043-2},
archivePrefix = {arXiv},
       eprint = {2510.25336},
 primaryClass = {astro-ph.SR},
       adsurl = {https://ui.adsabs.harvard.edu/abs/2025LRSP...22....4K}
}

@ARTICLE{Zhou2025,
       author = {{Zhou}, Yuhao},
        title = "{The formation of solar prominences: plasma origin and mechanisms}",
      journal = {Reviews of Modern Plasma Physics},
         year = 2025,
        month = dec,
       volume = {9},
       number = {1},
          eid = {32},
        pages = {32},
          doi = {10.1007/s41614-025-00206-6},
archivePrefix = {arXiv},
       eprint = {2511.14374},
 primaryClass = {astro-ph.SR},
       adsurl = {https://ui.adsabs.harvard.edu/abs/2025RvMPP...9...32Z}
}

@ARTICLE{vranjes2013,
       author = {{Vranjes}, J. and {Krstic}, P.~S.},
        title = "{Collisions, magnetization, and transport coefficients in the lower solar atmosphere}",
      journal = {\aap},
         year = 2013,
        month = jun,
       volume = {554},
          eid = {A22},
        pages = {A22},
          doi = {10.1051/0004-6361/201220738},
archivePrefix = {arXiv},
       eprint = {1304.4010},
 primaryClass = {astro-ph.SR},
       adsurl = {https://ui.adsabs.harvard.edu/abs/2013A&A...554A..22V}
}

@ARTICLE{Wargnier2022,
       author = {{Wargnier}, Q.~M. and {Mart{\'\i}nez-Sykora}, J. and {Hansteen}, V.~H. and {De Pontieu}, B.},
        title = "{Detailed Description of the Collision Frequency in the Solar Atmosphere}",
      journal = {\apj},
         year = 2022,
        month = jul,
       volume = {933},
       number = {2},
          eid = {205},
        pages = {205},
          doi = {10.3847/1538-4357/ac6e62},
       adsurl = {https://ui.adsabs.harvard.edu/abs/2022ApJ...933..205W}
}

@ARTICLE{gibson2018,
       author = {{Gibson}, Sarah E.},
        title = "{Solar prominences: theory and models. Fleshing out the magnetic skeleton}",
      journal = {Living Reviews in Solar Physics},
         year = 2018,
        month = dec,
       volume = {15},
       number = {1},
          eid = {7},
        pages = {7},
          doi = {10.1007/s41116-018-0016-2},
       adsurl = {https://ui.adsabs.harvard.edu/abs/2018LRSP...15....7G}
}

@ARTICLE{terradas2013,
       author = {{Terradas}, J. and {Soler}, R. and {D{\'\i}az}, A.~J. and {Oliver}, R. and {Ballester}, J.~L.},
        title = "{Magnetohydrodynamic Waves in Two-dimensional Prominences Embedded in Coronal Arcades}",
      journal = {\apj},
         year = 2013,
        month = nov,
       volume = {778},
       number = {1},
          eid = {49},
        pages = {49},
          doi = {10.1088/0004-637X/778/1/49},
archivePrefix = {arXiv},
       eprint = {1309.4934},
 primaryClass = {astro-ph.SR},
       adsurl = {https://ui.adsabs.harvard.edu/abs/2013ApJ...778...49T}
}

@ARTICLE{Bakhareva1992,
       author = {{Bakhareva}, N.~M. and {Zaitsev}, V.~V. and {Khodachenko}, M.~L.},
        title = "{Dynamic Regimes of Prominence Evolution}",
      journal = {\solphys},
         year = 1992,
        month = jun,
       volume = {139},
       number = {2},
        pages = {299-314},
          doi = {10.1007/BF00159156},
       adsurl = {https://ui.adsabs.harvard.edu/abs/1992SoPh..139..299B}
}

@ARTICLE{Stepanov2024,
       author = {{Stepanov}, A.~V. and {Zaitsev}, V.~V. and {Kupriyanova}, E.~G.},
        title = "{Features of the Joule Dissipation in the Solar Atmosphere}",
      journal = {Geomagnetism and Aeronomy},
         year = 2024,
        month = dec,
       volume = {64},
       number = {8},
        pages = {1203-1214},
          doi = {10.1134/S0016793224700300},
       adsurl = {https://ui.adsabs.harvard.edu/abs/2024Ge&Ae..64.1203S}
}

@ARTICLE{goodman2004,
       author = {{Goodman}, M.~L.},
        title = "{On the efficiency of plasma heating by Pedersen current dissipation from the photosphere to the lower corona}",
      journal = {\aap},
         year = 2004,
        month = mar,
       volume = {416},
        pages = {1159-1178},
          doi = {10.1051/0004-6361:20031719},
       adsurl = {https://ui.adsabs.harvard.edu/abs/2004A&A...416.1159G}
}

@ARTICLE{Heinzel2025,
       author = {{Heinzel}, Petr and {Beck}, Dominik and {Gun{\'a}r}, Stanislav and {Anzer}, Ulrich},
        title = "{Radiative Processes in Cool Coronal Condensations}",
      journal = {\solphys},
         year = 2025,
        month = nov,
       volume = {300},
       number = {12},
          eid = {166},
        pages = {166},
          doi = {10.1007/s11207-025-02569-y},
       adsurl = {https://ui.adsabs.harvard.edu/abs/2025SoPh..300..166H}
}

@ARTICLE{anzer2007,
       author = {{Anzer}, U. and {Heinzel}, P.},
        title = "{Is the magnetic field in quiescent prominences force-free?}",
      journal = {\aap},
         year = 2007,
        month = jun,
       volume = {467},
       number = {3},
        pages = {1285-1288},
          doi = {10.1051/0004-6361:20066817},
       adsurl = {https://ui.adsabs.harvard.edu/abs/2007A&A...467.1285A}
}

@ARTICLE{anzer2009,
       author = {{Anzer}, U.},
        title = "{Global prominence oscillations}",
      journal = {\aap},
         year = 2009,
        month = apr,
       volume = {497},
       number = {2},
        pages = {521-524},
          doi = {10.1051/0004-6361/200811107},
       adsurl = {https://ui.adsabs.harvard.edu/abs/2009A&A...497..521A}
}

@ARTICLE{hood1990,
       author = {{Hood}, A.~W. and {Anzer}, U.},
        title = "{A Model for Quiescent Solar Prominences with Normal Polarity}",
      journal = {\solphys},
         year = 1990,
        month = mar,
       volume = {126},
       number = {1},
        pages = {117-133},
          doi = {10.1007/BF00158302},
       adsurl = {https://ui.adsabs.harvard.edu/abs/1990SoPh..126..117H}
}

@ARTICLE{heinzel2024,
       author = {{Heinzel}, Petr and {Gun{\'a}r}, Stanislav and {Jej{\v{c}}i{\v{c}}}, Sonja},
        title = "{Partial ionization of plasma in solar prominences}",
      journal = {Philosophical Transactions of the Royal Society of London Series A},
         year = 2024,
        month = jun,
       volume = {382},
       number = {2272},
          eid = {20230221},
        pages = {20230221},
          doi = {10.1098/rsta.2023.0221},
       adsurl = {https://ui.adsabs.harvard.edu/abs/2024RSPTA.38230221H}
}

@ARTICLE{Karpen2001,
       author = {{Karpen}, J.~T. and {Antiochos}, S.~K. and {Hohensee}, M. and {Klimchuk}, J.~A. and {MacNeice}, P.~J.},
        title = "{Are Magnetic Dips Necessary for Prominence Formation?}",
      journal = {\apjl},
         year = 2001,
        month = may,
       volume = {553},
       number = {1},
        pages = {L85-L88},
          doi = {10.1086/320497},
       adsurl = {https://ui.adsabs.harvard.edu/abs/2001ApJ...553L..85K}
}

@ARTICLE{Hashimoto2023,
       author = {{Hashimoto}, Yuki and {Ichimoto}, Kiyoshi and {Huang}, Yuwei},
        title = "{Plasma diagnostics and Alfv{\'e}n wave heating of solar prominences by multiwavelength observations}",
      journal = {\pasj},
         year = 2023,
        month = oct,
       volume = {75},
       number = {5},
        pages = {913-924},
          doi = {10.1093/pasj/psad049},
       adsurl = {https://ui.adsabs.harvard.edu/abs/2023PASJ...75..913H}
}

@INPROCEEDINGS{karpen2015,
       author = {{Karpen}, Judith T.},
        title = "{Plasma Structure and Dynamics}",
    booktitle = {Solar Prominences},
         year = 2015,
       editor = {{Vial}, Jean-Claude and {Engvold}, Oddbj{\o}rn},
       series = {Astrophysics and Space Science Library},
       volume = {415},
        month = jan,
        pages = {237},
          doi = {10.1007/978-3-319-10416-4_10},
       adsurl = {https://ui.adsabs.harvard.edu/abs/2015ASSL..415..237K}
}

@ARTICLE{Kuperus1974,
       author = {{Kuperus}, M. and {Raadu}, M.~A.},
        title = "{The Support of Prominences Formed in Neutral Sheets}",
      journal = {\aap},
         year = 1974,
        month = mar,
       volume = {31},
        pages = {189},
       adsurl = {https://ui.adsabs.harvard.edu/abs/1974A&A....31..189K}
}

@INCOLLECTION{arregui2024,
       author = {{Arregui}, I{\~n}igo and {Van Doorsselaere}, Tom},
        title = "{Coronal heating}",
    booktitle = {Magnetohydrodynamic Processes in Solar Plasmas},
         year = 2024,
       editor = {{Srivastava}, Abhishek Kumar and {Goossens}, Marcel and {Arregui}, Inigo},
        pages = {415-450},
          doi = {10.1016/B978-0-32-395664-2.00015-3},
       adsurl = {https://ui.adsabs.harvard.edu/abs/2024mpsp.book..415A}
}

@ARTICLE{pecseli2000,
       author = {{P{\'e}cseli}, Hans and {Engvold}, Oddbj{\O}rn},
        title = "{Modeling of prominence threads in magnetic fields: Levitation by incompressible MHD waves}",
      journal = {\solphys},
         year = 2000,
        month = may,
       volume = {194},
       number = {1},
        pages = {73-86},
          doi = {10.1023/A:1005242609261},
       adsurl = {https://ui.adsabs.harvard.edu/abs/2000SoPh..194...73P}
}

@ARTICLE{parenti2007,
       author = {{Parenti}, S. and {Vial}, J.-C.},
        title = "{Prominence and quiet-Sun plasma parameters derived from FUV spectral emission}",
      journal = {\aap},
         year = 2007,
        month = jul,
       volume = {469},
       number = {3},
        pages = {1109-1115},
          doi = {10.1051/0004-6361:20077196},
       adsurl = {https://ui.adsabs.harvard.edu/abs/2007A&A...469.1109P}
}

@ARTICLE{low1975,
       author = {{Low}, B.~C.},
        title = "{Nonisothermal magnetostatic equilibria in a uniform gravity field. II. Sheet models of quiescent prominences.}",
      journal = {\apj},
         year = 1975,
        month = may,
       volume = {198},
        pages = {211-217},
          doi = {10.1086/153594},
       adsurl = {https://ui.adsabs.harvard.edu/abs/1975ApJ...198..211L}
}

@ARTICLE{sykora2017,
       author = {{Mart{\'\i}nez-Sykora}, Juan and {De Pontieu}, Bart and {Carlsson}, Mats and {Hansteen}, Viggo H. and {N{\'o}brega-Siverio}, Daniel and {Gudiksen}, Boris V.},
        title = "{Two-dimensional Radiative Magnetohydrodynamic Simulations of Partial Ionization in the Chromosphere. II. Dynamics and Energetics of the Low Solar Atmosphere}",
      journal = {\apj},
         year = 2017,
        month = sep,
       volume = {847},
       number = {1},
          eid = {36},
        pages = {36},
          doi = {10.3847/1538-4357/aa8866},
archivePrefix = {arXiv},
       eprint = {1708.06781},
 primaryClass = {astro-ph.SR},
       adsurl = {https://ui.adsabs.harvard.edu/abs/2017ApJ...847...36M}
}

@ARTICLE{Blokland2011,
       author = {{Blokland}, J.~W.~S. and {Keppens}, R.},
        title = "{Toward detailed prominence seismology. I. Computing accurate 2.5D magnetohydrodynamic equilibria}",
      journal = {\aap},
         year = 2011,
        month = aug,
       volume = {532},
          eid = {A93},
        pages = {A93},
          doi = {10.1051/0004-6361/201117013},
archivePrefix = {arXiv},
       eprint = {1106.4933},
 primaryClass = {astro-ph.SR},
       adsurl = {https://ui.adsabs.harvard.edu/abs/2011A&A...532A..93B}
}

@ARTICLE{lin2005,
       author = {{Lin}, Yong and {Wiik}, Jun Elin and {Engvold}, Oddbj{\o}rn and {Rouppe Van Der Voort}, Luc and {Frank}, Zoe A.},
        title = "{Solar Filaments and Photospheric Network}",
      journal = {\solphys},
         year = 2005,
        month = apr,
       volume = {227},
       number = {2},
        pages = {283-297},
          doi = {10.1007/s11207-005-1111-9},
       adsurl = {https://ui.adsabs.harvard.edu/abs/2005SoPh..227..283L}
}

@ARTICLE{arregui2018,
       author = {{Arregui}, I{\~n}igo and {Oliver}, Ram{\'o}n and {Ballester}, Jos{\'e} Luis},
        title = "{Prominence oscillations}",
      journal = {Living Reviews in Solar Physics},
         year = 2018,
        month = dec,
       volume = {15},
       number = {1},
          eid = {3},
        pages = {3},
          doi = {10.1007/s41116-018-0012-6},
       adsurl = {https://ui.adsabs.harvard.edu/abs/2018LRSP...15....3A}
}

@ARTICLE{low1981,
       author = {{Low}, B.~C. and {Wu}, S.~T.},
        title = "{A class of analytic solutions for the thermally balanced magnetostatic prominence sheet}",
      journal = {\apj},
         year = 1981,
        month = aug,
       volume = {248},
        pages = {335-343},
          doi = {10.1086/159158},
       adsurl = {https://ui.adsabs.harvard.edu/abs/1981ApJ...248..335L}
}

@ARTICLE{schmitt1995,
       author = {{Schmitt}, D. and {Degenhardt}, U.},
        title = "{Equilibrium and Stability of Quiescent Prominences.}",
      journal = {Reviews in Modern Astronomy},
         year = 1995,
        month = jan,
       volume = {8},
        pages = {61-80},
       adsurl = {https://ui.adsabs.harvard.edu/abs/1995RvMA....8...61S}
}

@ARTICLE{xia2011,
       author = {{Xia}, C. and {Chen}, P.~F. and {Keppens}, R. and {van Marle}, A.~J.},
        title = "{Formation of Solar Filaments by Steady and Nonsteady Chromospheric Heating}",
      journal = {\apj},
         year = 2011,
        month = aug,
       volume = {737},
       number = {1},
          eid = {27},
        pages = {27},
          doi = {10.1088/0004-637X/737/1/27},
archivePrefix = {arXiv},
       eprint = {1106.0094},
 primaryClass = {astro-ph.SR},
       adsurl = {https://ui.adsabs.harvard.edu/abs/2011ApJ...737...27X}
}

@ARTICLE{donne2024,
       author = {{Donn{\'e}}, D. and {Keppens}, R.},
        title = "{Mass Cycle and Dynamics of a Virtual Quiescent Prominence}",
      journal = {\apj},
         year = 2024,
        month = aug,
       volume = {971},
       number = {1},
          eid = {90},
        pages = {90},
          doi = {10.3847/1538-4357/ad50a3},
archivePrefix = {arXiv},
       eprint = {2405.20048},
 primaryClass = {astro-ph.SR},
       adsurl = {https://ui.adsabs.harvard.edu/abs/2024ApJ...971...90D}
}

@ARTICLE{popescu2025,
       author = {{Popescu Braileanu}, B. and {Keppens}, R.},
        title = "{Coronal rain formation in a two-fluid approximation}",
      journal = {\aap},
         year = 2025,
        month = jun,
       volume = {698},
          eid = {A189},
        pages = {A189},
          doi = {10.1051/0004-6361/202554712},
archivePrefix = {arXiv},
       eprint = {2505.03930},
 primaryClass = {astro-ph.SR},
       adsurl = {https://ui.adsabs.harvard.edu/abs/2025A&A...698A.189P}
}

@ARTICLE{gilbert2007,
       author = {{Gilbert}, Holly and {Kilper}, Gary and {Alexander}, David},
        title = "{Observational Evidence Supporting Cross-field Diffusion of Neutral Material in Solar Filaments}",
      journal = {\apj},
         year = 2007,
        month = dec,
       volume = {671},
       number = {1},
        pages = {978-989},
          doi = {10.1086/522884},
       adsurl = {https://ui.adsabs.harvard.edu/abs/2007ApJ...671..978G}
}

@ARTICLE{tsinganos1992,
       author = {{Tsinganos}, K. and {Surlantzis}, G.},
        title = "{MHD equilibria with flows in uniform gravity. I - 1-D prominence- and arcade-type solutions}",
      journal = {\aap},
         year = 1992,
        month = jun,
       volume = {259},
       number = {2},
        pages = {585-594},
       adsurl = {https://ui.adsabs.harvard.edu/abs/1992A&A...259..585T}
}

\appendix

\section{Convergence of the solutions}

\label{sec:resolution}
\begin{figure}
    \centering
    \includegraphics[width=0.9\linewidth]{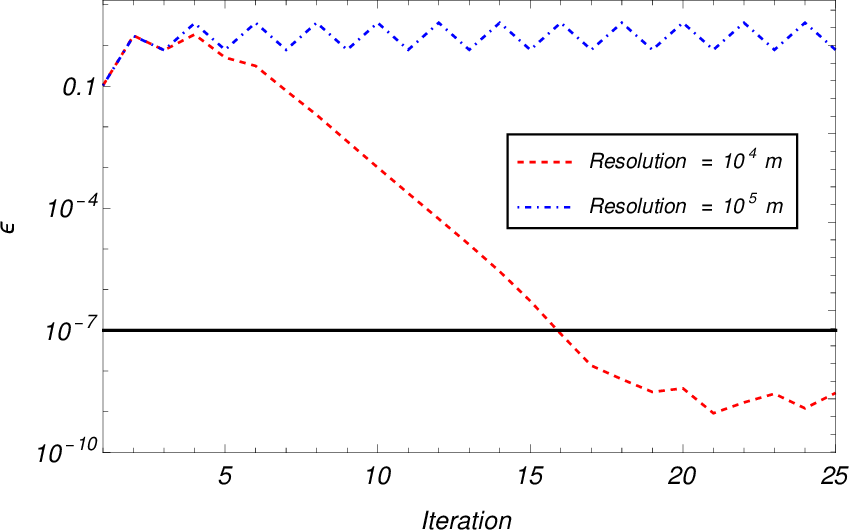}
    \caption{Comparison of the convergence of the parameter $\varepsilon$ from iteration to iteration for two cases with different spatial resolutions. In both cases, the model parameters  are $T_{0}=8000$~K, $p_{0}=5$~mPa, $B_{0}=10$~G, and $\phi=45^{\circ}$. The horizontal line denotes the convergence threshold of $\varepsilon = 10^{-7}$.}
    \label{fig:convergence}
\end{figure}

Here, we briefly discuss the converge of the self-consistent method to compute equilibrium solutions. First, we explore the effect of the spatial resolution used in the numerical integration the energy balance equation to compute the temperature profile. Figure \ref{fig:convergence} shows the evolution  of the convergence parameter $\varepsilon$, as defined in Eq.~(\ref{eq:epsilon}), from iteration to iteration, for two different spatial resolutions and considering the same model parameters. For this set-up, only the case with high resolution correctly converges to a self-consistent model, as $\varepsilon$ progressively decreases until the  convergence threshold of $\varepsilon = 10^{-7}$ is reached. We note that this threshold is arbitrary and is set after examining the evolution of many converging calculations.   High resolution is crucial in regions where strong gradients are present. In this problem, the PCTR exhibits sharp variations in several physical quantities, which  require a very fine resolution to be properly resolved. In the case with low resolution, the PCTR is not sufficiently resolved. Instead of converging to a self-consistent model, the iterative method enters a loop in which two different states alternate indefinitely. The convergence properties of the method can be compared with what was already discussed in \cite{melis2023heat}. In the previous paper, depending on the set of model parameters, the method either converged and produced cool thread models or became unstable, preventing the computation of physically-consistent models.

\begin{figure}
    \centering
    \includegraphics[width=0.9\linewidth]{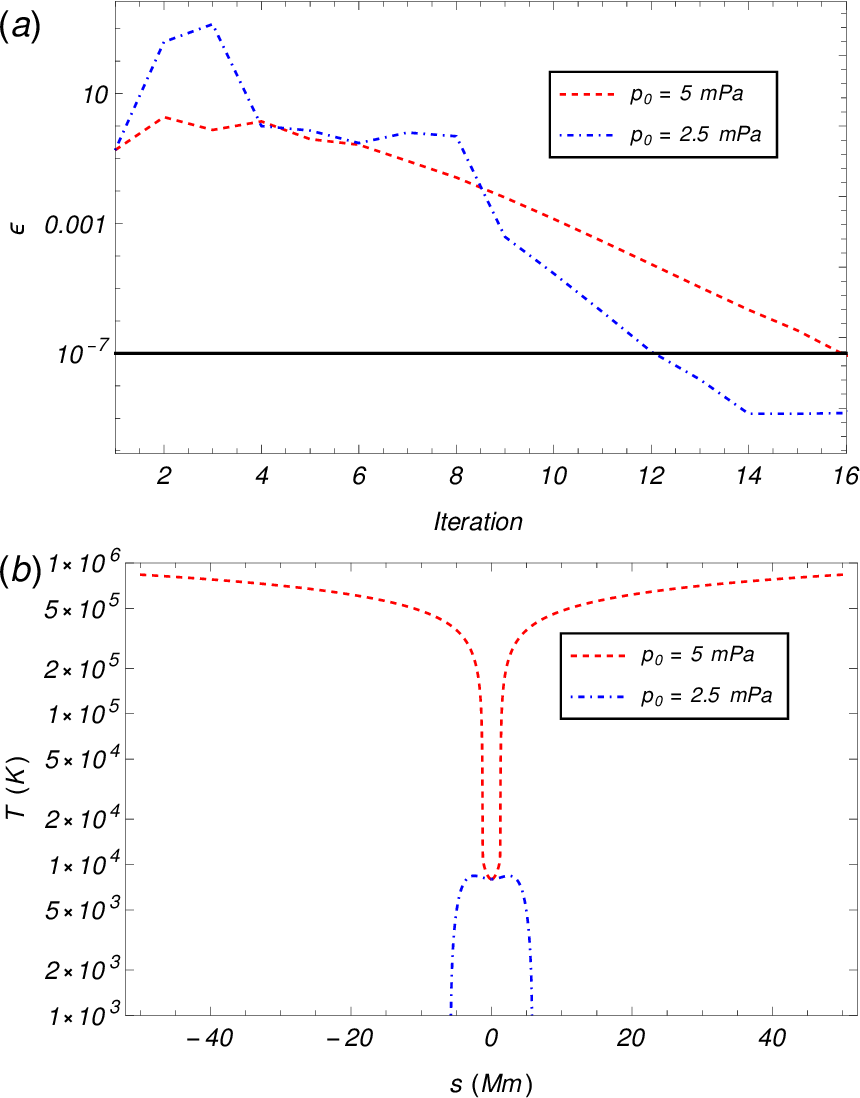}
    \caption{(a) Comparison of the convergence of the parameter $\varepsilon$ from iteration to iteration for two cases with different central pressures. The horizontal line denotes the convergence threshold of $\varepsilon = 10^{-7}$. (b) Converged temperature profiles for the same cases. In both cases, the model parameters  are $T_{0}=8000$~K,  $B_{0}=10$~G, and $\phi=88^{\circ}$.}
    \label{fig:loop}
\end{figure}

Another constraint that heavily affects the computation of physically acceptable models is  the condition in Equation~(\ref{eq:constraint}), which corresponds to the  requirement that the radiative losses at the thread center must be larger than the heating. In cases in which this condition is not satisfied, it may still be possible to compute models that apparently converge according to the criterion based on the value of $\varepsilon$. However, those models are nonphysical. Figure \ref{fig:loop}(a) shows a comparison of the evolution of $\varepsilon$ in two situations: a case in which Equation~(\ref{eq:constraint}) is satisfied and another case in which it is not. Essentially, the difference between the two cases is the assumed value of the central pressure, $p_0$. We find that  both cases converge, with convergence being slightly faster in the low-pressure case than in the high-pressure case. Although both cases apparently converge, the computed temperature profiles differ significantly. This is shown in Figure~\ref{fig:loop}(b), where the final temperature profiles of both cases are compared. The high-pressure case satisfies Equation~(\ref{eq:constraint}) and the computed temperature profile has the expected shape. Conversely, Equation (\ref{eq:constraint}) is not satisfied in the low-pressure computation, resulting in a temperature profile that displays an unexpected behavior. Instead of increasing toward coronal values when $|s|$ increases, the temperature decreases and eventually reaches negative values, which is physically impossible. The reason for this behavior is that, when ambipolar heating  exceeds  radiative losses at the center,  thermal conduction must become a negative term in the cool region to compensate the excess of heating. Consequently, the developed temperature gradient has the opposite sign (decreasing temperatures with increasing $|s|$) than in the physically-acceptable solutions. Obviously, although the method has converged, these unacceptable solutions must be discarded.

\section{Comparison with \citet{milne1979prominences}.}
\label{sec:beta}

Our research shares many similarities with the work of \citet{milne1979prominences}, and also with the more recent paper by \citet{vigh2018model}, although there are several important differences.  Here, we discuss how these previous results can be compared with our findings.

\citet{milne1979prominences} assumed a fully ionized plasma, but we considered   partial ionization, which incorporates two effects that have an important influence in the energy balance:  ambipolar diffusion  and the role of neutrals in the thermal conduction. Another relevant difference resides in the boundary conditions used to solve the energy balance equation. In our work, we imposed the temperature and the pressure at the center of the thread, in order to have a dense and cold region, and the energy equation is integrated from the center to the edge of the domain as in an initial-value problem. Conversely,  \citet{milne1979prominences} set the temperature and the density at the edge of the domain to coronal values, while they imposed  symmetry of both the vertical component of the magnetic field and  the temperature at the center. Thus, in the case of \citet{milne1979prominences}, the energy equation is integrated as a two-point, boundary-value problem, where the central temperature and pressure have to be determined rather than being fixed. 

Because of the different numerical set-up, our results are not directly equivalent to those of \citet{milne1979prominences}, although some comparisons can be made. \citet{milne1979prominences} discussed the existence of two types of solutions, the cool solutions and the hot solutions, whose appearance crucially depended on the plasma $\beta = p_0/(B_0^2/2\mu_0)$, defined as the ratio between the gas and magnetic pressures, and the shear angle. Only in the case of the cool solutions, the central temperature is low enough to be consistent with prominence temperatures.   Essentially, the temperature, pressure and density profiles in the cool solutions of \citet{milne1979prominences} are similar to the ones computed in the present paper, having a cold and dense region at the center and corona-like regions in the outermost part of the domain, as is seen   in Figs.~10 and 11 of \citet{milne1979prominences}. Hot solutions are not possible in our method, since we already imposed the temperature at the center to be that of cold prominences.

\begin{figure}
    \centering
    \includegraphics[width=\linewidth]{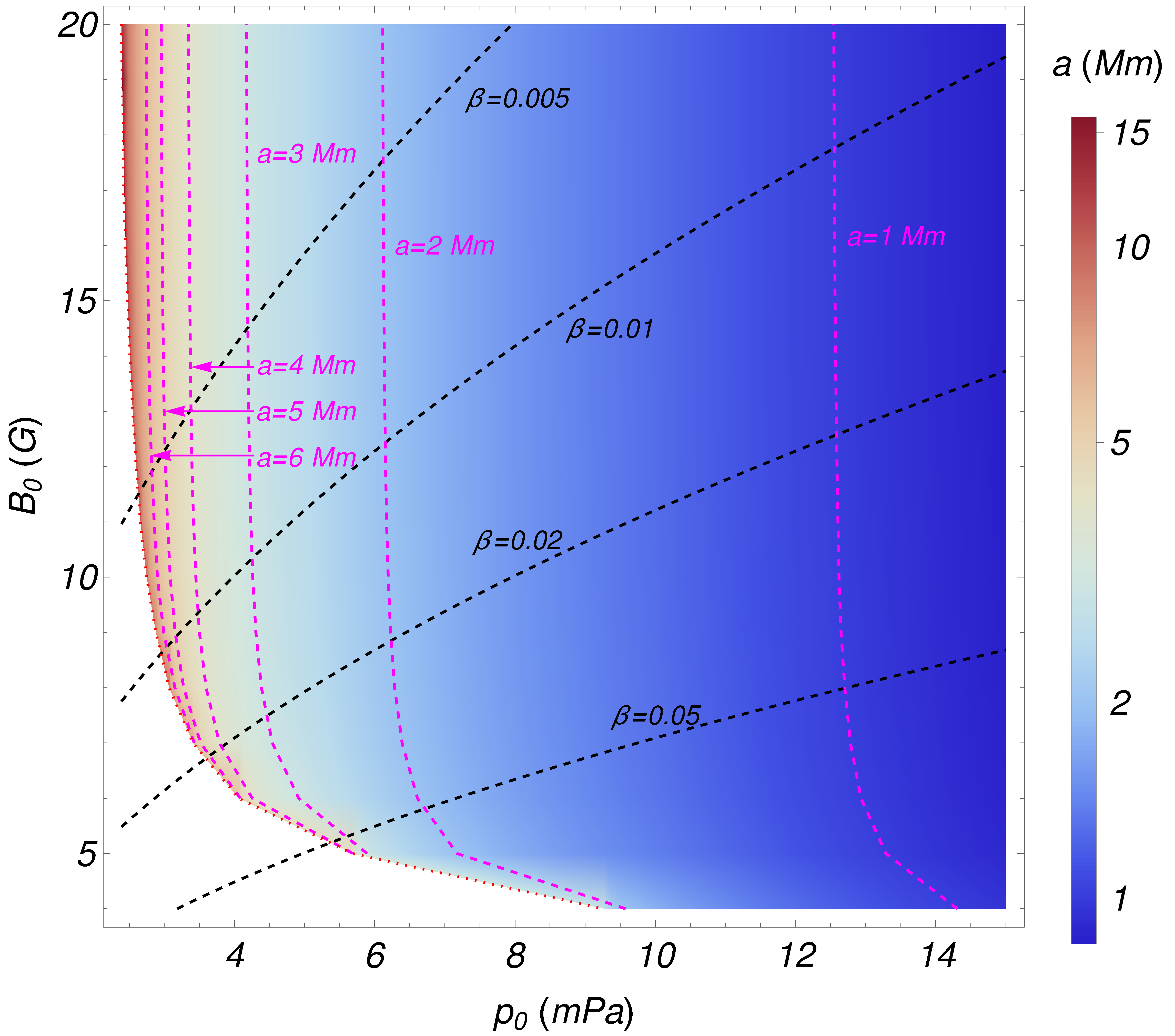}
    \caption{Contour plot of the cold thread length, $a$, as a function of $p_{0}$ and $B_{0}$. The black dashed lines represent contours of constant $\beta$, while the purple dashed lines represent contours of constant thread length. All the models are computed with $T_{0}=$~8,000~K and $\phi=88^{\circ}$. }
    \label{fig:beta}
\end{figure}

\citet{milne1979prominences}  explained that there was a small range of $\beta$ for which a cool equilibrium was computable. However, such allowed values of $\beta$ are  higher than the ones considered here. To further address this, Figure~\ref{fig:beta} displays a contour plot of the cold thread length as a function of both $p_{0}$ and $B_{0}$, for $T_{0}=$~8,000~K and $\phi=88^{\circ}$. The  dashed purple lines denote several combinations of $p_{0}$ and $B_{0}$ that provide the same cold thread length (indicated next to every line).  As discussed  in Sect.~\ref{sec:param}, equilibrium models with high pressures have short cold threads, almost regardless of the magnetic field strength. Conversely, for low pressures, the computed thread lengths  can reach values above 10~Mm for  intense  magnetic fields. Cold and dense threads with realistic lengths can only exist for sufficiently intense magnetic fields  and relatively low pressures. This is explained by the fact that the weight of the dense plasma must be balanced by the upward magnetic  force. If the magnetic field is too weak, the dense thread cannot be sustained.   For every value of $p_0$, there is a critical value of $B_0$ above which an equilibrium is possible, and we see that the cold thread length increases abruptly when $B_0$ approaches the critical value. The domain where  an equilibrium can be reached is enclosed by a red dotted line in Figure \ref{fig:beta}, while no equilibrium is allowed in the white region. The black dashed lines included in Figure~\ref{fig:beta} indicate constant contours of $\beta$. Regions with relatively large $\beta$ are related with short cold threads and with limited ranges of $p_0$ and $B_0$ where an equilibrium is possible. On the contrary, regions with small $\beta$ are associated with longer cold threads and with wider ranges for $p_0$ and $B_0$. However, it is clear from the results of Figure~\ref{fig:beta} that the plasma $\beta$ does not uniquely determine the length of the cold thread in our models. In fact, models with dramatically different cold thread lengths are possible along contours of constant $\beta$.

The values of the cold thread length obtained in these computations are far larger than the ones obtained in \cite{milne1979prominences} for their standard model, but they obtained longer cold regions for larger values of the shear angle. This particular result agrees with our  findings, although in the case of \cite{milne1979prominences} there was a maximum shear angle for an equilibrium to exist due to the constraints imposed by their boundary conditions. Such a maximum shear does not appear in our models. We recall that, in our models, the plasma ionization degree is also important and determines several physical parameters, such as the thermal conductivity and the ambipolar diffusion coefficient. The assumed value of $T_0$ is very relevant concerning the  plasma ionization degree.  Therefore, the properties of our models and, in particular, the length of the cold thread, results from a complicated interplay between the four parameters: $p_0$, $B_0$, $T_0$, and $\phi$. Conversely,   $\beta$ and $\phi$ are the two crucial parameters that fully determine the properties of the models computed by  \cite{milne1979prominences}.

\end{document}